\documentclass{aa}  

\usepackage{graphicx}
\usepackage{txfonts}
\usepackage{xcolor}
\usepackage{amssymb}
\usepackage[T1]{fontenc}
\usepackage[normalem]{ulem}

\begin{document}

\title{Galactoseismology in cosmological simulations}
\subtitle{Vertical perturbations by dark matter, satellite galaxies and gas}

% The list of authors, and the short list which is used in the headers.
% If you need two or more lines of authors, add an extra line using \newauthor

   \author{B. García-Conde\inst{1}
          \and
          T. Antoja\inst{2,3,4}
          \and
          S. Roca-Fàbrega \inst{5}
          \and
          F. Gómez \inst{6,7}
            \and
          P. Ramos \inst{8}
            \and
          N. Garavito-Camargo \inst{9}
          \and
          MA. Gómez-Flechoso \inst{1, 10}
          }

\institute{Dpto. de Física de la Tierra y Astrofísica, Fac. CC. Físicas, Universidad Complutense de Madrid, Plaza de las Ciencias, 1, E-28040 Madrid, Spain.
              \email{begona01@ucm.es}
\and
{Institut de Ci\`encies del Cosmos (ICCUB), Universitat de Barcelona (UB), c. Mart\'i i Franqu\`es, 1, 08028 Barcelona, Spain}
\and 
{Departament de Física Qu\`antica i Astrof\'isica (FQA), Universitat de Barcelona (UB),  c. Mart\'i i Franqu\`es, 1, 08028 Barcelona, Spain}
\and 
{Institut d'Estudis Espacials de Catalunya (IEEC), c. Gran Capit\`a, 2-4, E-08034 Barcelona, Spain}
\and
{Lund Observatory, Division of Astrophysics, Department of Physics, Lund University, Box 43, SE-221 00 Lund, Sweden.}
\and
{Instituto Multidisciplinar de Investigaci\'on  y Postgrado, Universidad de La Serena, Ra\'ul Bitr\'an 1305, La Serena, Chile.}
\and
{Departamento de Astronom\'ia, Universidad de La Serena, Av. Juan Cisternas 1200 Norte, La Serena, Chile.}
\and
{National Astronomical Observatory of Japan, Mitaka-shi, Tokyo 181-8588, Japan.}
\and
{Center for Computational Astrophysics, Flatiron Institute, 162 Fifth Ave, New York, NY 10010, USA. }
\and
{Instituto de Física de Partículas y del Cosmos (IPARCOS), Universidad Complutense de Madrid, E-28040 Madrid, Spain.}
}
% These dates will be filled out by the publisher
\date{Accepted 2023 November 10. Received 2023 November 10; in original form 2023 July 13}


 \abstract
{%Context
Only recently, complex models that include the global dynamics from dwarf satellite galaxies, dark matter halo structure, gas infalls, and stellar disk in a cosmological context became available to study the dynamics of disk galaxies such as the Milky Way (MW). }
%Aims
{We use a MW model from a suite of high-resolution hydrodynamical cosmological simulations named GARROTXA to establish the relationship between the vertical disturbances seen in its galactic disk and multiple perturbations, from the dark matter halo, satellites and gas. }
%Methods
{We calculate the bending modes in the galactic disk in the last  6~Gyr of evolution. To quantify the impact of dark matter and gas we compute the vertical acceleration exerted by these components onto the disk and compare them with the bending behavior with Fourier analysis.}
%Results
{We find complex bending patterns at different radii and times, such as an inner retrograde mode with high frequency, as well as an outer slower retrograde mode excited at different times. The amplitudes of these bending modes are highest during the early stages of the thin disk formation (20 km s$^{-1}$) and reach up to 8.5 km s$^{-1}$ in the late disk evolution. We find that the infall of satellite galaxies leads to a tilt of the disk, and produces strong anisotropic gas accretion with a misalignment of 8$^{\circ}$ with subsequent star formation events, and supernovae, creating significant vertical accelerations onto the disk plane. The misalignment between the disk and the inner stellar/dark matter triaxial structure, formed during the ancient assembly of the galaxy, also creates a strong vertical acceleration on the stars. We also find dark matter sub-halos that temporally coincide with the appearance of bending waves in certain periods. }
%Conclusions
{We conclude that several agents trigger the bending of the stellar disk and its phase spirals in this simulation, including satellite galaxies, dark sub-halos, misaligned gaseous structures, and the inner dark matter profile. These phenomena coexist and influence each other, sometimes making it challenging to establish direct causality.
}


\keywords{Galaxy: kinematics and dynamics  -- galaxies: evolution  -- methods: numerical --  stars: kinematics }
\maketitle

%%%%%%%%%%%%%%%%%%%%%%%%%%%%%%%%%%%%%%%%%%%%%%%%%%

%%%%%%%%%%%%%%%%% BODY OF PAPER %%%%%%%%%%%%%%%%%%

\section{Introduction}\label{sec:intro}

The stars of the Milky Way's (MW) disk appear to be vertically disturbed. These perturbations can be detected in the stellar density distribution of structures like the warp, corrugations, and north-south asymmetries  \citep{widrow2012galactoseismology,williams2013wobbly, schonrich2018warp}. More recently, using the high-quality data of stars in the Solar Neighbourhood from the Gaia satellite \citep{gaia2018gaia, collaboration2021gaia}, \citet{antoja18} found a new structure in the phase-space: the phase spiral. This structure is the result of a perturbation to the disk (external or internal) which vertically displaces stars and triggers a phase mixing process. To explain its origin, the focus has been put on the Sagittarius dwarf galaxy, which is a known recent perturber of the Milky Way \citep{binney2018origin, laporte2019footprints, bland2019galah, Li2020, Bland-Hawthorn2021, hunt2021resolving, gandhi2021snails, widmark2021weighing}. Recent observations have shown that the formation mechanism of the phase spiral is more complex than initially expected \citep{hunt2022multiple, antoja2022phase, alinder2023investigating, frankel2023vertical}. These structures in the vertical phase space are products of the perturbations of the stellar disk, which can be seen as deformations in both the kinematics and positions with respect to the galactic plane. These deformations can also be detected as bending waves. This bending behavior has also been detected in other galaxies \citep{Gomez2021,nandakumar2022bending, urrejola2022winds}.

Several theoretical studies using N-body and analytical models have attributed vertical perturbations to interactions with satellites, \citep[e.g.][]{Gomez2013, debattista2014vertical, de2015phase,d2016excitation, Laporte2018, semczuk2020tidally, Bennett2021}, the dark matter wake as a consequence of the interaction between the halo and an external perturber \citep{Weinberg1998, gomez2016fully,laporte2018b, grand2022dark}, and the interaction with dark matter sub-halos \citep{feldmann2015detecting, Chequers2018, tremaine2023origin}. Other amplification mechanisms of the vertical perturbations of the disk include gas infalls along warps \citep{ khachaturyants2022bending} and misalignment of the plane of stars at different radii \citep{sellwood2022internally}. %driving the subsequent evolution of vertical perturbations 
Thus, vertically disturbed stars can have a diverse range of origins and a perturbed disk may be a common state throughout galactic disks.

\cite{grion2021holistic} used a purely gravitational N-body simulation to calculate the accelerations produced by the disk, the Sagittarius-like dwarf galaxy, and dark matter halo to discern the dominant source of perturbations in the different parts of the disk and different moments of evolution. They found that there are moments when the dark matter halo's response to the passing of the satellite enhances its effect. However, to have a complete picture of the dynamical mechanisms taking place in the Galaxy, one should include the hydrodynamical processes. Cosmological hydrodynamical simulations take into account realistic satellite accretion into the host galaxy and hence multiple satellites can perturb the disk, whereas isolated N-body models only account for one perturbing satellite interacting with the disk. However, multiple satellites can lead to different effects when interacting with the disk and with each other \citep{rocha2012infall,widrow2014bending, d2016excitation, trelles2022concurrent}. In \citet{gomez2017} they used cosmological simulations to study vertical patterns in galactic disks. They found that approximately  70\% of the simulated galaxies show clear vertical patterns produced by close satellite encounters, distant flybys of massive companions, accretion of misaligned cold gas, and re-accretion of cold gas from the progenitors of gas-rich major mergers. However, the spatial resolution of the simulation does not allow to study scales of the size of the solar neighborhood. The study of \citet{van2015creation} used a cosmological model to follow the evolution of highly misaligned gas, where the inner parts of the galaxy re-align with it faster, resulting in a warp at later times of evolution. \citet{semczuk2020tidally} used galactic disks from the IllustrisTNG cosmological simulation to study tidally induced warps, finding that both satellite interaction and gas accretion may induce warps.

In our previous work (\citealt{garcia2022phase}; GC22 hereafter), we found phase spirals in a Milky Way-like disk for the first time in a cosmological simulation, namely the GARROTXA \citep{garrotxa} model. However, its exact origin was not clear, since the satellites in this model are less massive ($\sim 10^{9}$ $M_{\odot}$ at the moment of the pericenter) than the minimum mass reported in the bibliography to excite phase spirals, at least in the context of the Sagittarius dwarf galaxy and the Milky Way. For example, the models by \citet{laporte2019footprints} and \citet{Bland-Hawthorn2021} used a mass of $6 \times 10^{10}$  and $2 \times 10^{10}$ M$_{\odot}$ respectively. We suggested that other effects such as the interaction of the collective effects of several satellites, or the misalignment of the angular momentum of the different components of the simulation, especially the stellar disk and dark matter halo, or the combination of some of them can generate strong perturbations in the disk. Later on, \cite{grand2022dark} used a high-resolution simulation of a Milky Way-like galaxy from the Auriga suite of cosmological simulations and resolved phase spirals in the disk. The main formation mechanism of the phase spiral in that simulation was found to be the dark matter halo response to the passage of a satellite galaxy. Differently from our case, they have a dominant source of perturbation, so they can identify the main cause of the triggering of the phase spiral.

Here, we further analyze the vertical behavior of the disk of the GARROTXA model. In this work, we identify bending modes (as calculated in \citealt{widrow2014bending}) and we study their behavior using Fourier decomposition. We calculate the vertical acceleration of the dark matter particles onto the disk and its relation with the passing satellites and the inner structure of the halo itself. We also analyze the vertical gravitational acceleration applied by the gas to understand the relationship between its misalignment with respect to the disk. Thus, we present a more realistic picture of the satellites, halos, gas, and disk interactions. 

In this paper, we start in Sec. \ref{sec:garrotxa} by describing the simulation and its different structures. In Sec. \ref{sec:disk} we describe the behavior of the disk and quantify its disturbances, focusing on the vertical component. In Sec. \ref{sec:accelerations} we calculate the vertical accelerations ($a_{Z}$) applied by dark matter and gas onto the disk and identify all satellites in the system. In Sec. \ref{sec:results} we link the measured bending modes and disk tilting to the different typology of interactions occurring in our model, characterized in the previous sections, and we find many intertwined agents involved in the bending of the disk. In the last section, we discuss the results and we present our conclusions.

\section{The GARROTXA Simulation}\label{sec:garrotxa}

In this work, we use model G.323 of the GARROTXA simulation \citep{garrotxa}, which is similar (although at the lower end of the predicted virial mass range) to the Milky Way in terms of disk size and rotation curve. The spatial resolution of this model is 109 pc, with a minimum dark matter particle mass of $10^5$ $M_{\odot}$, a typical mass of disk stellar particles of $3.5 \times 10^{3}$ $M_{\odot}$, and a minimum time-step of $10^3$ yr. At $z \sim 0$, we have $\sim 2 \times 10^6$ stellar particles within the virial radius. The summary of these characteristics is shown in Table \ref{table:characteristics}.

The analysis presented here focuses on data encompassing the last 6.3 Gyr (from 6.3 Gyr in lookback time to the present, equivalent to scale factors of $a=0.6$ to $a=1.0$, respectively) of the evolution of the simulated galactic system. This is the period where we find a well-defined stellar disk. In this period, we have 190 snapshots available. This allows us to undertake an exhaustive temporal evolution study of the disk’s kinematics and density structures. To read and analyze the simulation we use the software \textit{yt}  \citep{yt}. 

  \begin{table}
      \caption[]{Main characteristics of the MW-sized model G.323 of GARROTXA simulation at z$=0$.  From left to right, virial radius, virial mass, the total  stellar particles mass, mass of the disk, the typical mass of the disk particles, and number of disk particles. }
         \label{table:characteristics}
      $$
         \begin{array}{llllll}
            \hline
            \noalign{\smallskip}
               R_{\mathrm{vir}}  &M_{\mathrm{vir}} & M_{\ast} & M_{\mathrm{disk}}&M_{\mathrm{part}, \mathrm{disk}} & N_{\mathrm{part}, \mathrm{disk}}\\
                \noalign{\smallskip}
               [\mathrm{kpc}] & [M_{\odot}] &  [M_{\odot}] & [M_{\odot}] &[M_{\odot}] & \\
            \noalign{\smallskip}
            \hline
            \noalign{\smallskip}
            154 & 6.5 \times 10^{11}  &6\times 10^{10}& 2.6 \times 10^{10}& 10^{3}-10^{4} & 2\times10^6 \\

            \noalign{\smallskip}
            \hline
         \end{array}
     $$
   \end{table}

The main dark matter structures of this model at the mentioned period are a halo with a virial radius $R_{\mathrm{vir}}=154$ kpc 
\footnote{As calculated in GC22, computed following \citet{bryan1998statistical} where, they use the spherical collapse model to determine the virial overdensity $\Delta_{\rm{v}}(z)$  as a function of the redshift, and taking this value to be 333, i.e., $R_{\rm{333}}$=$R_{\rm{v}}$, hereafter} with a total mass of $6.5$ $\times 10^{11}$ $M_{\odot}$. The inner parts of the dark matter halo are dominated by a central triaxial substructure with an ellipsoidal shape extending to approximately 10 kpc in its major axis and 5 kpc in its minor axis. This structure was formed at z$\sim$ 3 (11.6 Gyr in lookback time) as a result of a 1:1 major merger \citep{garrotxa}. The oldest stellar particles, both belonging to the primordial galaxies (pre-merger) and born during this major merger, show properties similar to an old elliptical galaxy embedded in a tri-axial dark matter halo, which is not rotationally supported. After this 1:1 major merger, a new stellar disk starts to form but is perturbed by another merger, which is the last major merger of the galactic system studied here. This last major merger occurs at z$\sim$1.5 ($\sim$1:10 merger at 9.55 Gyr of lookback time) and sets the more recent components of the structure of the galaxy that is a thick-old stellar disk and a much younger thin stellar disk. 

%\s{The last major merger that the galaxy suffers at z$\sim$ 1.5 ($\sim$1:10 merger at 9.55 Gyr of lookback time). After the first merger, the galactic system develops a stellar thick disk structure and after the last one, a thinner stellar disk is formed and grows in radius. At z$\sim 0$, the disk has a mass of $2.6$ $\times10^{10}$ $M_{\odot}$. }

At each snapshot, we apply a centering and alignment of the main galactic system. The center of mass is calculated iteratively, in spheres of shrinking radii, as in \cite{power2003inner}. After this centering, we compute the disk plane by calculating the angular momentum of stars inside a 20\% of $R_{\mathrm{vir}}$, and we then apply a change of basis to align this with the $Z$ axis. This places most of the galactic disk in a plane oriented with $Z=0$. The X-axis is the projection of the original X-axis from the cosmological box onto the galactic plane whose normal vector is $L$. Then, we take all stellar particles born after z$\sim$ 1.5 inside a fixed cylindrical section with $Z<|3|$ and  $R<7$ kpc and apply the same process  which further aligns the inner galaxy's $L$ with the $Z$ axis.  %Despite this careful process, due to the complexity of cosmological disks, different parts of the disc show a slight misalignment of their angular momentum by 1-2 degrees with the adopted references system (Fig. \ref{fig:Lstars_vs_Lgas_ref_system} in Appendix \ref{sec:misalignment_appendix}). Overall, the intermediate parts (10-15 kpc) are the better aligned ones. }
We use galactocentric cylindrical coordinates $(R, \phi, Z)$ where $\phi$ and $V_{\phi}$ are negative in the direction of rotation. The direction of $\phi=0$ is oriented towards the positive $X$ axis. %vector $(1,0)$.

\begin{figure}
	\includegraphics[width=1\columnwidth]{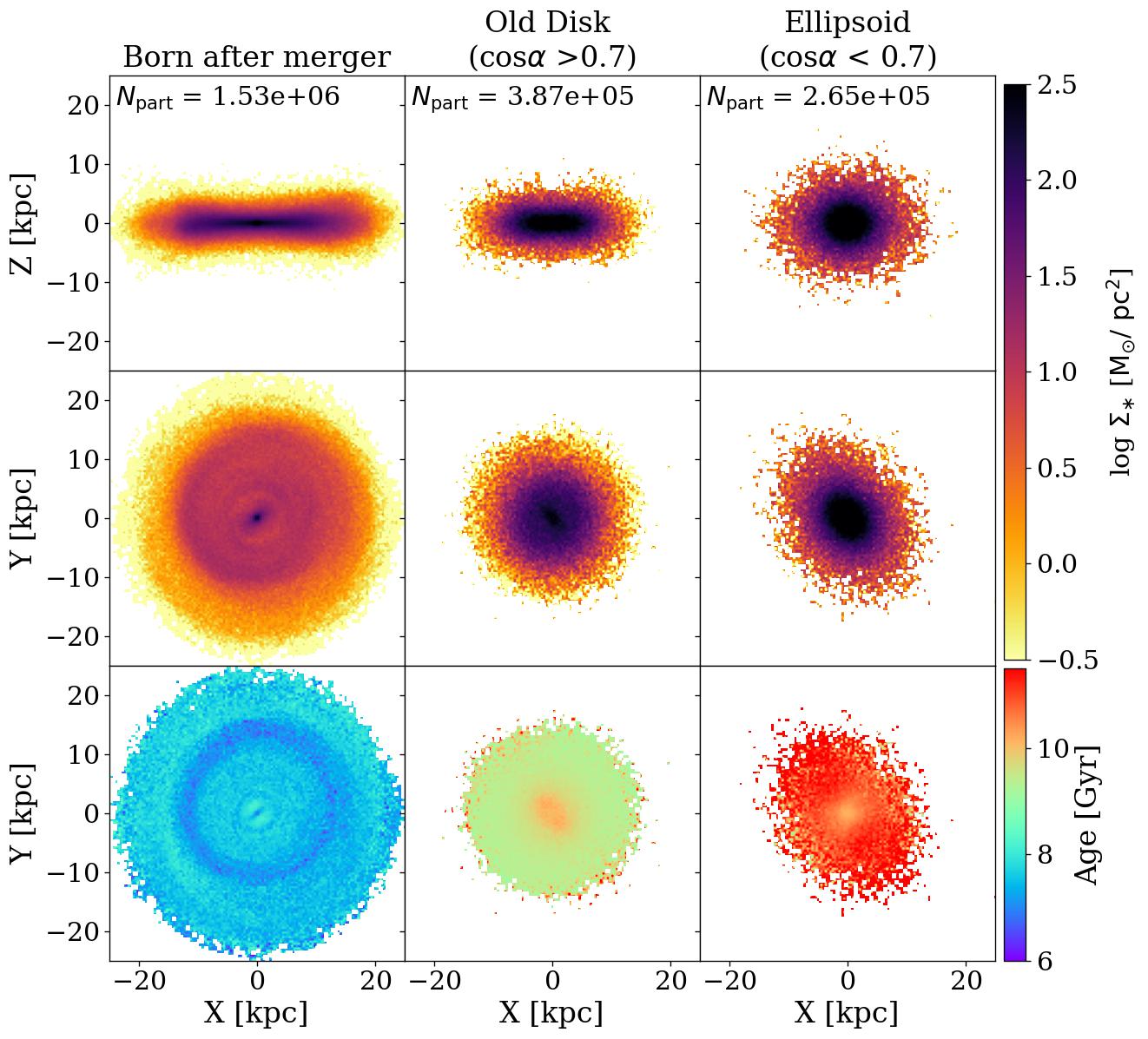}
    \caption{Different structures of the stellar component of the snapshot at present times $z=0$. The left panel shows stars born 500 Myr after the last major merger (at 9.55 Gyr in lookback time). The panels in the central column show the older, hot disk. Lastly, in the right column, we show the old stellar ellipsoid. The top two rows show the surface density of the stellar particles in the X-Z  and X-Y projections, respectively. The last row shows the face-on projection colored by age. The disk rotates clockwise in this reference system.}
    \label{fig:disk_nondisk_example}
\end{figure}

In Fig.~\ref{fig:disk_nondisk_example} we show the different structures of the stellar component of the model at the present time ($z=0$) in its edge-on and face-on projections (top and central rows, respectively) and the face-on view colored by age (bottom panels). We separate stellar particles depending on whether they were born after (left column) or before the last major merger mentioned above (middle and right columns). In practice, we take a time of 9.05 Gyr (i.e. 500 Myr after the merger) to ensure that we are selecting the particles born in the thin-young stellar disk.  

Regarding the stellar particles born before the merger, they are distributed in the above-mentioned ellipsoid (right column) akin to the one in dark matter, and a structure that today composes an old, dynamically hot disk (central column). To separate this old disk from the ellipsoid, we calculate the angle $\alpha$ between the vector of $Z=0$ plane and the angular momentum of the stars, considering stars with $\cos(\alpha)<0.7$ as part of the ellipsoid.

For our study, we use only particles born after the last major merger, thus belonging to the 'young' disk (left panels) to track any signs of vertical perturbations. In the inner galactic regions ($<$5~kpc), there is an oval structure whose major axis is oriented at the same angle at all times and has an $A_{2}/A_{0}$ average amplitude of 0.35. This elongated structure has a quadrupole signal in the velocity $V_{R}$, but since we do not detect a pattern speed, we do not consider it to be a bar. This structure contains stellar particles of the same age as the disk. The semi-major axis of this young elongated structure is oriented as the old-stellar and dark matter inner triaxial structure's minor axis. There is also a one-armed structure, or lopsided mode, in the young disk stellar component. We come back to this aspect in Appendix~\ref{sec:density_breathing_relation}, although for this study we focus mostly on the vertical disturbances. Finally, for the following analyses, we further select stellar particles with $|Z|< 2.5$ kpc and $R<25$ kpc, similar to GC22. In summary, in this study, we select particles from the disk based on their age (born after 9 Gyr in lookback time) and their location (aforementioned region). After this selection, we end up with $\sim1.39 \times 10^6$ stellar particles in the disk.

\section {Vertical bending modes of the disk} \label{sec:disk}

In this section, we focus on the $Z$ and $V_{Z}$ distributions of the disk through its bending modes, similarly to \citet{widrow2014bending}, as a way to quantify the disk's vertical deformation. These modes can lead to phase spirals \citep{Bland-Hawthorn2021, Darling2019}, and therefore our additional motivation is to be able to explain those found in GC22. By identifying the timing and spatial location of the perturbations, we hope to establish a link with the underlying phenomenon, whether internal or external, that cause this deformation. 

To calculate the vertical modes, we divide the X-Y plane into 70 bins from -25 to 25 kpc in both coordinates X and Y (bin size of $\approx 0.7$ kpc). This binning allows us to have enough stellar particles (an average of 350 particles per bin) to perform the following calculations. In each bin, we calculate the linear regression of $V_{Z}$ as a function of $Z$. The slope and the intercept correspond to the amplitudes of the breathing (A) and bending (B) amplitude respectively:

\begin{equation}
    V_{Z}(x,y;z)= A(x,y) z + B (x,y)
	\label{eq:bending_breathing_regression}
\end{equation}

%The bending (middle panels) and breathing (bottom panels) amplitudes of the  young disk  computed with Eq1 are shown in Fig 2 for lookback times 2.33  Gyr (left columns) and 0.99 Gyr (right columns).  

\begin{figure}
	\includegraphics[width=1\columnwidth]{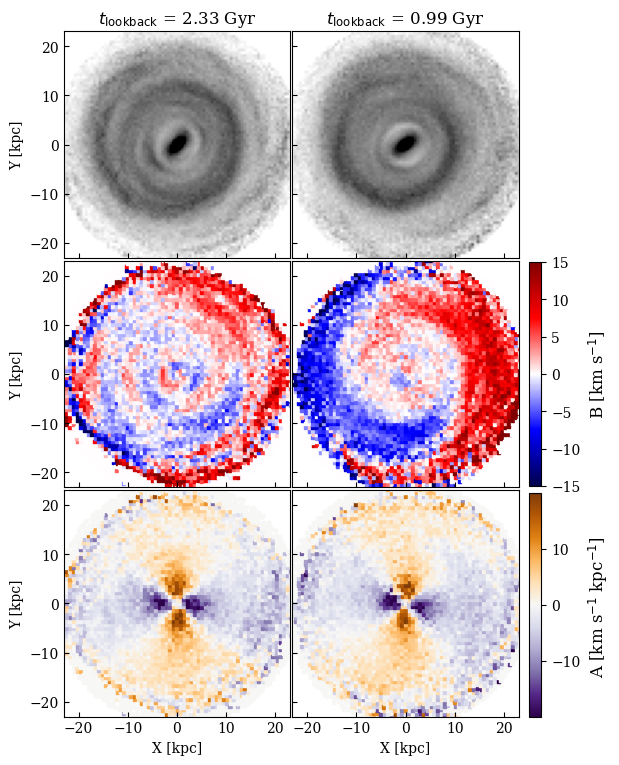}
    \caption{Overdensity plot of the disk} particles with a Gaussian filter $\sigma=6$ (top panel), bending (middle), and breathing (bottom) maps at two different snapshots of this simulation in each column, 2.33 Gyr and 0.99 Gyr in lookback time. A strong bending mode can be seen at 0.99 Gyr (middle right panel).
    \label{fig:density_bending_breathing_example}
\end{figure}

An example of this method can be seen in Fig.~\ref{fig:density_bending_breathing_example}, where we show the face-on projection of the galactic disk at two different lookback times of evolution corresponding to 2.33 Gyr (left columns) and 0.99  Gyr (right columns). We show an overdensity plot of the disk particles with a Gaussian filter (top panels), the bending (middle panels) and breathing (bottom panels) amplitudes of the young disk computed with Eq. \ref{eq:bending_breathing_regression}. In the top panels, we can see a one-armed density structure, which is leading and retrograde. In the central panels, we show how strong bending waves appear at specific times while they can be weaker in others (right vs left panels). In the breathing modes (bottom panels), we see a prominent mode $m=2$, which follows the oval structure at the inner parts of the stellar disk, and it is present all through the recent galaxy evolution. This is consistent with previous studies, where this breathing behavior is linked to the bar or other non-axisymmetric internal structures \citep{widrow2014bending, faure2014radial,monari2015vertical}.
 
Here we mostly focus on the bending modes, while the breathing modes and their relation to the density patterns (i.e. the central oval structure and the retrograde one-armed pattern) are briefly examined in Appendix ~\ref{sec:density_breathing_relation} for the sake of completeness. 

To visualize the evolution of the bending parameters over time, we take the values of $B$ in the $X$-$Y$ grid and apply a Fourier analysis that we evaluate in a polar grid, dividing the disk into radial sections with 1 kpc of width. For each ring, we have $N$ bins from the initial $X$-$Y$ grid, each of which is located in an azimuthal position $\phi_{i}$, and has a bending amplitude (weight in the Fourier decomposition) of $W_{i}=B_i$. Then, we perform a Fourier decomposition following the next expression\footnote{Similarly, we can analyze the breathing modes by doing $W_{i}=A_{i}$ and the density modes by using $W_{i}=M_{i}$. That is indeed what we do in Appendix ~\ref{sec:density_breathing_relation})}:

\begin{equation}
\begin{aligned}
    a_{m}=2 \sum_{i}^{N}{W_{i}\cos{(m \phi_{i})}} \\
    b_{m}=2 \sum_{i}^{N}{W_{i}\sin{(m \phi_{i})}} \\
    A_{m}=\sqrt{a_{m}^2 + b_{m}^2}/N
	\label{eq:fourier}
\end{aligned}
\end{equation}

We estimated the amplitudes for each mode, $A_{m}$, at each radius and time. We focus on the results for $m=1$ and $m=2$ in the rest of the analysis. We also analyzed higher modes, but, even though at some times they are comparable with the amplitudes of modes $m=1$ and $m=2$, they are never dominant. We also study the phase at which the maximum amplitude for each mode is located at each radius, following the expression:

\begin{equation}
%\begin{aligned}
    \phi_{m} = \mathrm{arctan} (b_{m}/a_{m}) \\
	\label{eq:fourier_phase}
%\end{aligned}
\end{equation}

\begin{figure*}
	\includegraphics[width=1\textwidth]{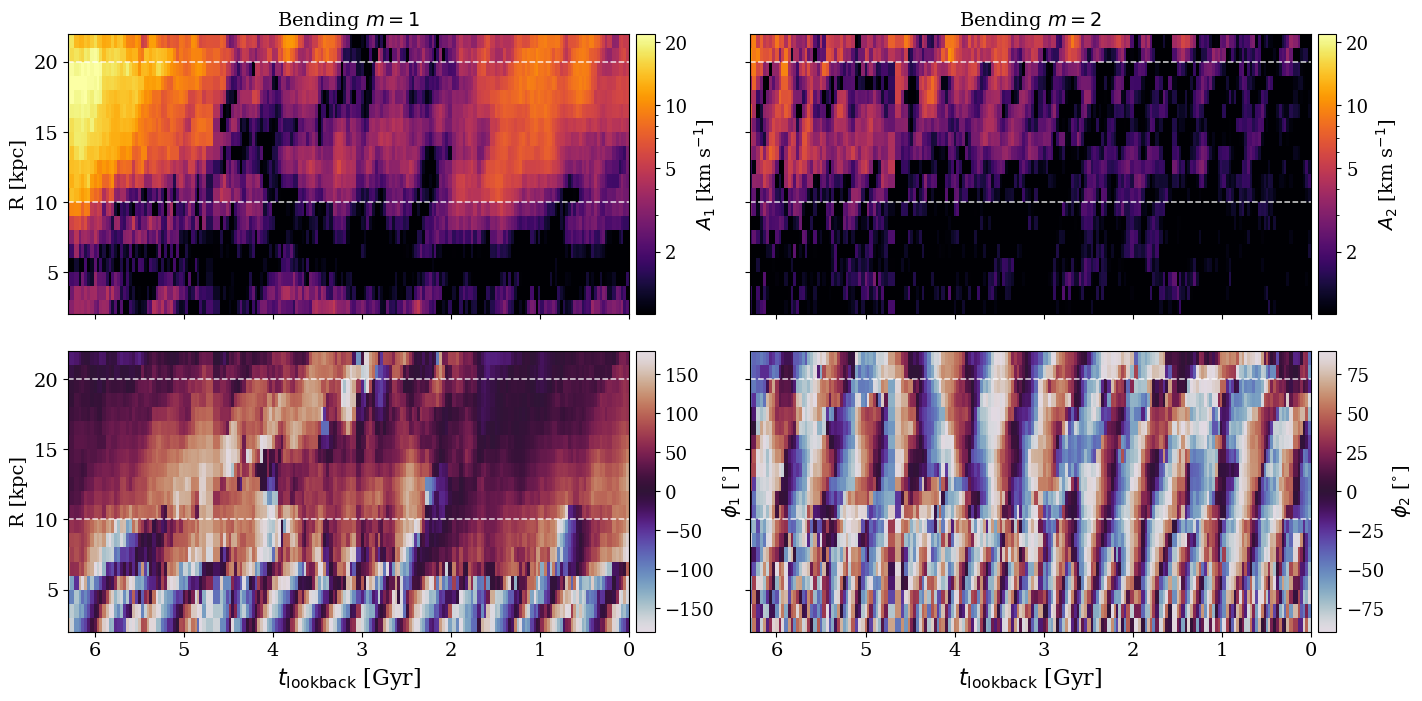}
    \caption{
    Temporal evolution of the bending modes' Fourier amplitude $A_{m}$ (top panels) and phase $\phi_{m}$ (bottom panels) of the $m=1$ (left panels) and $m=2$ (right panels) modes. The phase of mode $m=2$ ranges between $-90^{\circ}$ and $90^{\circ}$ due to the duplicity of the structures. The white horizontal grey-dashed lines mark the division of the disk at a radius of 10 and 20 kpc. The bending $m=1$ dominates over $m=2$ and shows higher amplitudes at outer radii (R$>7$ kpc) excited mainly at the beginning of the timespan studied and at later times from 2 Gyr on. The inner parts show faster retrograde bending waves with the highest amplitude happening around 4 Gyr of lookback time.}
    \label{fig:Bending_fourier}
\end{figure*}

In Fig.~\ref{fig:Bending_fourier} we show the amplitude and phases of the bending modes $m=1$ (left panels) and $m=2$ (right panels) as a function of radius and time. The bending pattern seems to be very complex.
Over the timespan studied, at $R>7$, there is an overall dominance of red-black colors (corresponding to a phase of 0-100$^{\circ}$), indicating that the bending is prone to be excited at this angle.
At the beginning of the evolution (between 6.3 and 5 Gyr), there is a highly perturbed disk, as indicated by the higher amplitudes of the bending modes (top row of panels), which can be up to 20 km s$^{-1}$ at external radii. At these times, the phase of this bending mode $m=1$ (bottom left panel) at radii $R>7$ shows a slow retrograde pattern ($\dot{\phi}>0$ in our reference system). 
%Over the timespan studied, there is an overall dominance of red-black colours (corresponding to a phase of 0-100 $^{\circ}$) at $R>7$, indicating that the bending is prone to be excited at this angle.
%This deformation of the disk seems to be present throughout all the time intervals studied but with different intensities: there is an overall dominance of red-black colours (corresponding to a phase of 0-100 $^{\circ}$) at these radii. 
%Observing its amplitude, it is excited at different times, such as from 2 Gyr on.
From 2 Gyr to the present, we see that the amplitude of the bending mode $m=1$ increases, with the phase oriented as in previous times, reaching its maximum at approximately 1.1 Gyr, with a value of 7.5 km s$^{-1}$ between 10 and 15 kpc, and 8.5 km s$^{-1}$ between 15 and 20 kpc with a leading shape.  %More specifically, the outer parts show a slow retrograde mode ($\dot{\phi}>0$ in our reference system).

At inner radii ($<$ 7 kpc), we generally see a fast retrograde pattern ($\dot{\phi}>0$, in the bottom left panel) that is excited at certain times (top left panel). We observe that its amplitude is maximum at 3.5 Gyr (with 3.5 km s$^{-1}$ of amplitude).

%\s{from 4.2 Gyr on
%we see an increase of the amplitude of the bending mode $m=1$ (top left panel) at the inner parts ($<$ 7 kpc), which reaches a maximum at 3.5 Gyr (with 4.5 km s$^{-1}$ of amplitude) and decreases until 3 Gyr. Its phase (bottom left panel) shows that it corresponds to a retrograde pattern ($\dot{\phi}>0$). From 2 Gyr to the present, we see that the amplitude of the bending mode $m=1$ increases, with the phase oriented as in previous times, suggesting that this deformation is being again excited. At this times, in radii larger than 10 kpc we see a slow retrograde bending wave, and its amplitude reaches its maximum at approximately 1.1 Gyr, with a value of 3 km s$^{-1}$ between 10 and 15 kpc, and 7.5 km s$^{-1}$ between 15 and 20 kpc. In the inner parts at $<$ 10 kpc, reaches its maximum at 1 Gyr with an amplitude of 3.5 km s$^{-1}$.}

The $m=2$ mode of the bending shows high amplitudes at earlier times (top left panel), and generally at higher radii, has prograde motion ($\dot{\phi}<0$). From 2.5 Gyr to 1.8 Gyr we see an increase at intermediate regions of the disk, although the bending mode $m=1$ generally dominates at these times and radii (we further discuss the behavior of the bending modes amplitudes in Sec.~\ref{sec:discussion}).

\section{Perturbation agents: Dark matter, satellite galaxies and gas} \label{sec:accelerations}

To quantify the impact that the dark matter and gas have on the disk's vertical direction, we have calculated the vertical gravitational accelerations $a_{Z}$ they produce on the galactic disk which is on the $Z=0$ plane at every snapshot (see Sec.~\ref{sec:garrotxa}). This is computationally faster than calculating accelerations on all disk particles. This mesh has 100 $\times$ 100 bins in the $X$ and $Y$ directions and extends to $R = 40$ kpc. We have used the following expression:

\begin{equation}
\begin{aligned}
    a_{z, bin}= \sum_{i}^{N_{part}}\frac {-G m_{i} }{r_{i,bin}^2} \frac{r_{z; i,bin}}{r_{i,bin}},\\
	\label{eq:acceleration}
\end{aligned}
\end{equation}

%\begin{equation}
%\begin{aligned}
%    a_{z, bin}= \sum_{i}^{N_{part}}\frac {-G m_{i} }{r_{i,bin}^2} \;  \bf{\hat{r}}_{z},\\
%	\label{eq:acceleration}
%\end{aligned}
%\end{equation}

\noindent where the vertical acceleration at each bin ($a_{z,\mathrm{bin}}$) is the sum of all individual vertical accelerations from all the mass elements $N_{\mathrm{part}}$ (dark matter particles or gas cell)  of each component (dark matter and gas). $m_{i}$ is the mass of each mass element, and $r_{i,\mathrm{bin}}$ is the total distance between the bin and the mass element. The relation $r_{z; i,\mathrm{bin}}/r_{i,\mathrm{bin}}$ takes the vertical component of the acceleration. %\t{The vector $\bf{\hat{r}}_{z}$ takes the vertical component of the acceleration}.

Some mass elements can be very close to the $Z=0$ plane, producing large accelerations, and increasing the noise in the Fourier analysis. To mitigate this, we establish a limit of maximum acceleration in a bin, corresponding to the acceleration an element can apply if it is situated at a vertical distance of $\pm 54 $ pc to the plane, which is half of the minimum size of the gas cell (equivalent to the spatial resolution of the simulation: 109 pc). If an acceleration in a bin exceeds this limit, it is set to this maximum.

%\s{we vertically shift these mass elements to have at least $\pm 54 $ pc,} 
\begin{figure*}
\centering
	\includegraphics[width=1\textwidth]{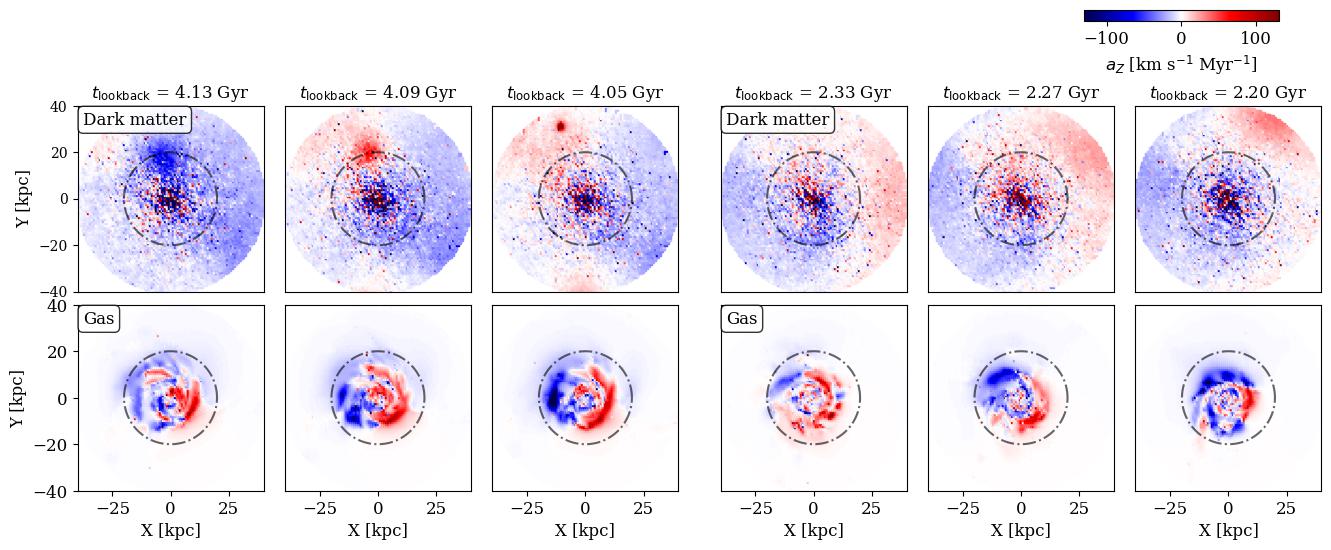}
    \caption{Vertical acceleration of dark matter (top panels) and gas (bottom panels). Black dot-dashed circles are situated at 20 kpc, illustrating the size of the stellar disk of the galaxy. The dark matter acceleration at the top three leftmost panels is dominated by the crossing of dark matter sub-halos DM1119 and DM2595. The effect of gas at these times (bottom panels) is bimodal. The dark matter's $a_{Z}$ of the right block of panels is dominated by the satellite galaxy "Escarabajo", which applies positive $a_{Z}$ in the positive x - positive y region of the mesh. }
    \label{fig:acceleration_example}
\end{figure*}

%We calculate the vertical acceleration applied by dark matter and gas in all snapshots. 
In Fig.~\ref{fig:acceleration_example} we show $a_{Z}$ in the $Z=0$ plane at two different time periods (three leftmost panels vs. the three rightmost) coinciding with two instances when there are at least one of the satellites near its pericentre. In the top row, we show the acceleration by the dark matter particles, where we can see the effects of the satellite, crossing the plane between the first panel (negative $Z$, negative vertical accelerations) and the next one (positive $Z$, positive vertical accelerations). The three rightmost panels show another interaction, but in this case, the satellite is not crossing the plane at the moment of minimum distance. However, we can see the vertical accelerations it induces in the outer parts of the grid (upper right regions in red), which are outside of the disk. 

In the bottom row of Fig.~\ref{fig:acceleration_example}, we show the acceleration exerted by the gas. We can see that it is stronger at the disk region, where it is more concentrated. At certain times (e.g. at 4.09 Gyr), the map appears to be divided into one half of the gaseous component inducing positive acceleration and the other, a negative one, suggesting the possibility of a tilted or warped gaseous disk. This will be further analyzed in Sec.~\ref{subsec:acceleration_gas}.

Once the calculation of $a_{Z}$ is carried out, we apply the same Fourier decomposition as in previous sections. This allows us to better visualize the evolution of the vertical accelerations with time, understand the effects that these have on the disk dynamics, and compare them with the results from the previous sections. In the next subsections, we present the accelerations from the different components.

\subsection{Accelerations by Dark Matter}
\label{sec:dm_accelerations}

\begin{figure*}
	\includegraphics[width=1\textwidth]{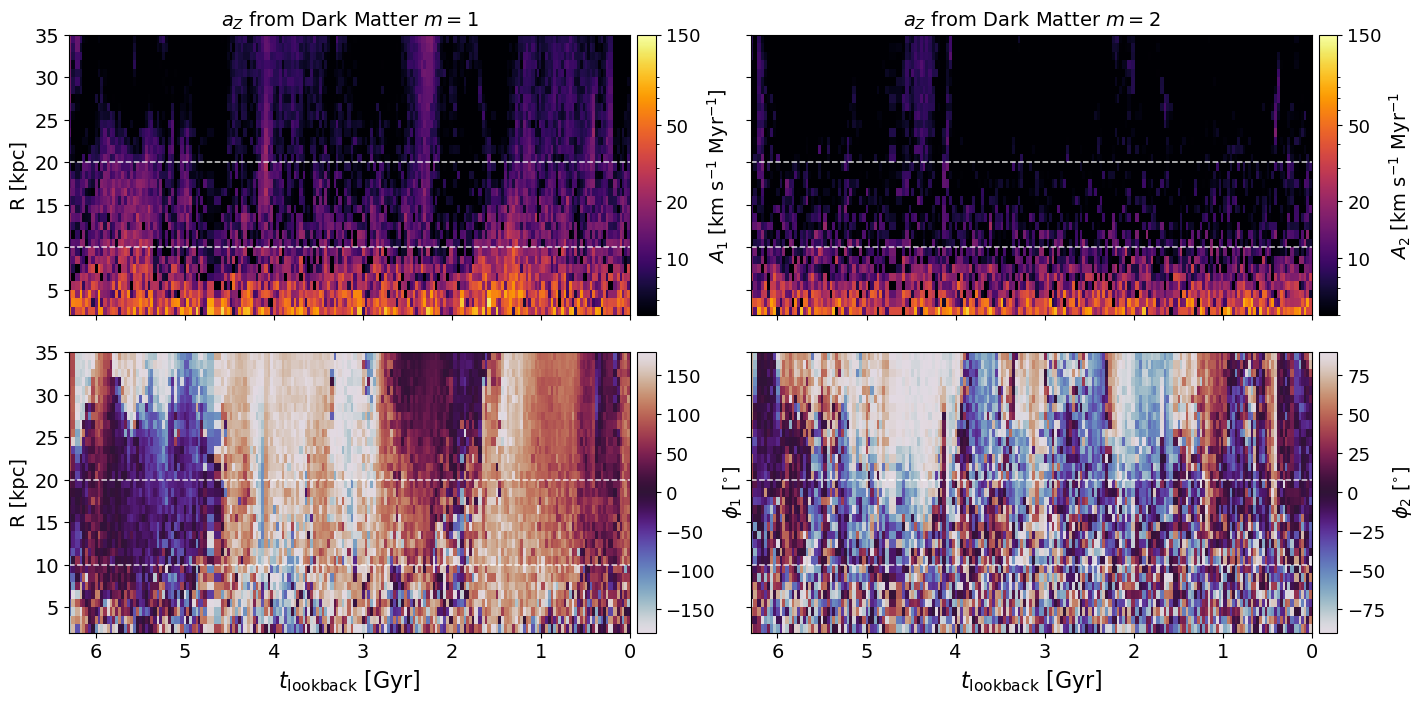}
    \caption{ Fourier decomposition of the vertical acceleration by dark matter particles onto the $Z=0$ plane. We show $m=1$ (left panels) and $m=2$, from which we have the amplitude of the mode (top panels) and the phase where the maximum amplitude is located (bottom panels). These are similar to \ref{fig:Bending_fourier}, but extended to 40 kpc of radius. The inner acceleration is dominated by the inner dark matter structure.}
    \label{fig:DM_acceleration_fourier}
\end{figure*}

Here we first take the approach of considering the impact of the dark matter as a whole, without excluding any particles belonging to satellites. This will include the effects of the global structure of the halo, its satellites, and associated unbound material, and the possible wakes excited by these satellites. The results are shown in Fig.~\ref{fig:DM_acceleration_fourier}, where we display the amplitude of the $m=1$ and $m=2$ modes and their respective phases of maximum amplitude, $\phi_{m}$. At inner radii ($<10$ kpc, and mostly in the $m=1$ mode), the particles that are likely to dominate the acceleration are the ones from the gravity of the inner structure of the dark matter halo by proximity, over the particles situated at higher radius, like the ones from the satellites. Indeed, in Sec.~\ref{sec:results} we show that the inner dark matter structure is triaxial and slightly tilted in relation to that of the disk, most of the time, producing a constant acceleration. This tilt is especially relevant at certain moments, such as 1.5 Gyr approximately, which coincides with the moments of higher accelerations at these radii. 

In the outer parts of the disk, the acceleration produced by the dark matter component ($>20$ kpc) comes mainly from the dark matter of satellites orbiting the system, especially when they cross the galactic plane. In the following Sec.~\ref{sec:satellites_rockstar} we study in detail the isolated effects of the satellites. 

\subsection{Accelerations by satellite dwarf galaxies and dark sub-halos}
\label{sec:satellites_rockstar}
\begin{figure*}
	\includegraphics[width=1\textwidth]{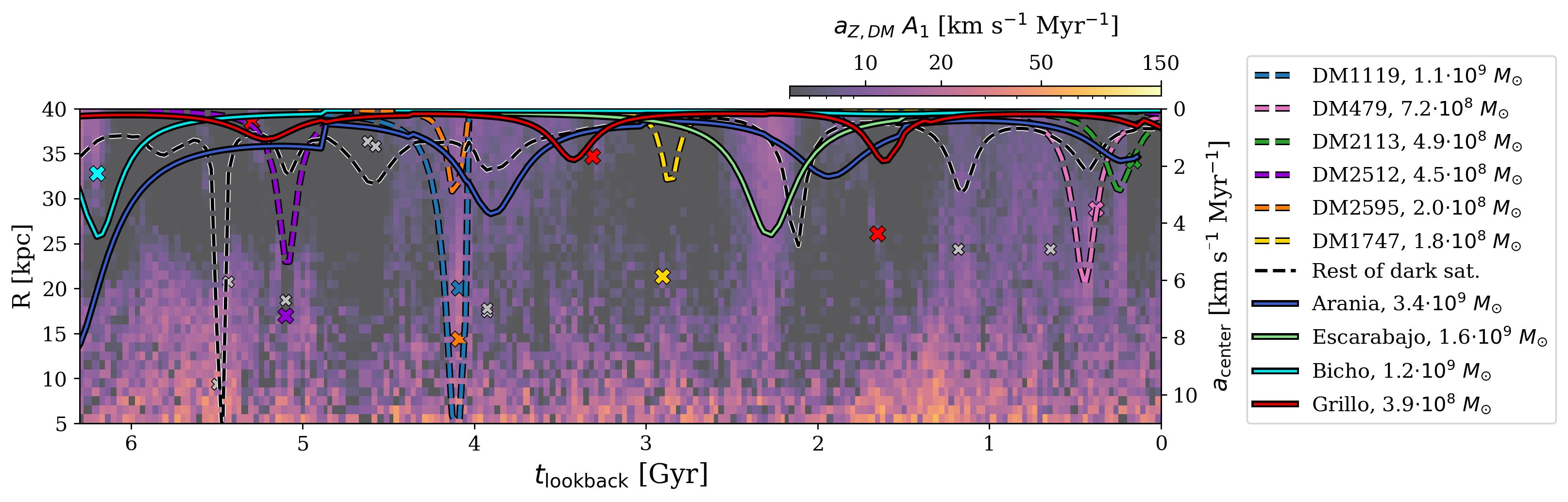}
    \caption{
    Total gravitational acceleration magnitude produced by each individual sub-halo on the center of the galactic system (colored lines, right y-axis) and the sum of acceleration by the rest of (less massive) sub-halos (black dashed line, right y-axis). Sub-halos containing stellar particles are indicated as solid lines, and those with pure dark matter are shown as dashed lines. We also marked the moments and positions (cylindrical radius, left axis) when the satellites ("x") cross the galactic plane. The grey "x" symbols correspond to the crossings of less massive satellites. Background color and left axis are equivalent to the top-left panel in Fig.~\ref{fig:DM_acceleration_fourier}. Notice that, while the background color shows the vertical acceleration by all dark matter on the galactic plane $Z=0$, lines show the total acceleration produced by sub-halos at the galactic center, not at the $Z=0$ plane. The latter gives proportionally lower values of acceleration due to the difference of the distance between the satellite and the closest point of the plane vs to the center of the galaxy. For the lines, we also use the data of $M_{\mathrm{bound}}$ for each satellite, re-evaluated at each apocenter. The masses indicated on the label are the first mass registered for each satellite, which corresponds to $M_{200}$. In this figure, we see that the vertical acceleration from the dark matter halo between 25-40 kpc is correlated with the pericentric passages of several satellites, dark or luminous. The ones exerting more vertical acceleration are DM1119 and Escarabajo, as well as DM479 and DM2113 at later times. } %We use colored "x" symbols for those dark matter halos with mass $>10^8$ $M_{\odot}$ (see legend) and grey "x" symbols for the less massive. }
    \label{fig:DM_vs_halos_rockstar}
\end{figure*}

The GARROTXA model is a zoom-in cosmological simulation, thus, the halo of the main galactic system contains many substructures both in stars and dark matter. Here we used the Rockstar halo finder \citep{behroozi2012rockstar} to identify all sub-halos that may impact the disk’s kinematics. With this halo finder, we identify all dark matter structures with $M_{200}$ $>$ $10^6$ $M_{\odot}$ (to avoid poorly resolved sub-halos and contamination by field halo dark matter particles), and, using the dark matter particle IDs we also find all gravitationally bound particles. If the sub-halos have stellar mass within 20\% of the $R_{200}$ of the satellite (as calculated by Rockstar) we identify them as luminous. We use the value of $M_{200}$ right before the infall into the $R_{Vir}$ of the host galaxy as a reference of the size and mass of the sub-halo. Once the satellites have entered the virial radius of the host, it is highly difficult to determine their mass. In these cases, we use or indicate the bound mass given by Rockstar.

The complete data of the sub-halos (including both dark and luminous satellites) found by Rockstar can be found in Fig.~\ref{fig:satellites_by_mass}, in the Appendix \ref{sec:satellites_appendix}. By applying this software, we have identified new dark satellites and some dwarf galaxies compared to GC22. In the timespan studied, we found nearly 500 independent sub-halos (luminous or dark) that enter the galaxy. As expected from the $\Lambda$CDM Universe, we find a large number of dark satellites, i.e. sub-halos with no stellar particles. For instance, inside the virial radius of the studied galaxy at redshift $z=0$, we find $139$ of the aforementioned sub-halos (those of lower mass get destroyed over time) without any stellar particle, compared to $10$ dwarf galaxies. These dark satellites, although they are generally less massive than dwarf galaxies, can interact with the disk as luminous satellites would. 

In Fig.~\ref{fig:DM_vs_halos_rockstar} we show again the top left panel of Fig.~\ref{fig:DM_acceleration_fourier}, i.e., the amplitude of $m=1$ mode of the dark matter vertical acceleration, but now showing also the total, not just vertical, acceleration from each sub-halo on the center of the galactic disk. The total acceleration by the most important dark satellites is shown with dashed lines, whereas the ones containing stellar particles are shown with solid lines. The radius and time at which each satellite crosses the plane are marked by a cross of the same color. We observe that the increase in the vertical acceleration in the regions from 25 to 40 kpc (background colors) is consistent with the close encounter with sub-halos both dark and luminous (lines and crosses). The constant disk crossing of the many dark and luminous satellites is permanently perturbing the disk. However, only those of higher mass, and those that get closer to the galactic plane show significant $a_{Z}$ (see background colors vs. lines). For instance, at 5.1 Gyr or 2.9 Gyr the dark satellites DM2512 and DM1747 (violet and yellow lines) cross the disk but we do not observe a strong increase in $a_{Z}$ (background colors).

%dark satellites DM1504 and DM2113 (ochre and dark green dashed lines, respectively) cross the disk. However, we do not observe a strong increase in $a_{Z}$ (background colors). 

In contrast, the strongest interactions that we observe are the following. First, a strong interaction occurs at 4.1 Gyr, and it is due to a dark satellite (DM1119, blue dashed curve) containing a mass of $1.1\times 10^{9}$ $M_{\odot}$. This interaction is fast, and the pericentre occurs when the object is crossing the plane, at 20 kpc. This satellite is accompanied by another dark satellite, DM2595, with a mass of  $10^{8}$ $M_{\odot}$ (orange), and the dwarf galaxy Arania’s second pericenter, whose mass at this moment is of $1.3 \times 10^{9}$ $M_{\odot}$ (blue line with stars).

Secondly, another noticeable moment is the interaction occurring at 2.2 Gyr with a dwarf galaxy which we will refer to as "Escarabajo" (light green). This satellite starts its interaction with $1.6\times 10^{9}$ $M_{\odot}$, being a slower interaction than the previous example at 4.1 Gyr. In this case, the pericenter does not occur near the galactic plane yet we see a band of strong accelerations in the background colors. The black dashed line is the sum of the accelerations of those satellites whose mass does not exceed $10^{8}$ $M_{\odot}$ or maximum acceleration does not exceed 2 km s$^{-1}$ Myr s$^{-1}$. This shows the accelerations of different dark matter sub-haloes that, even though they are less massive, may have a relation to some of the increases seen in the $a_{Z}$ applied by the dark matter particles. However, the increase in acceleration by dark matter is not always related to passing satellites. We find that at 1.1 Gyr, the increase is not entirely explained by the passing of minor satellites, as in the case of Fig.~\ref{fig:acceleration_example}. By examining the acceleration by dark matter in the X-Y projection, we find that this is related to the overall dark matter distribution. As will be discussed in Sec.~\ref{subsec:escarabajo}, at these times there are instabilities of the galaxy that reflect as a net acceleration onto the plane.  %This increase is also due to the instabilities of the galaxy (as will be discussed in the next sections), which can displace the center of mass between the galaxy and dark matter halo, thus reflecting as a net acceleration onto the plane.  
Finally, from around 1 Gyr until the present we observe several sub-halos, which overall increase the accelerations in the intermediate and outer regions, like the pink and light green ones with pericenters occurring at 0.4 and 0.26 Gyr respectively.
%\s{Interestingly, we observe that this sum of accelerations sometimes presents peaks, almost as intense as other individual satellites. For instance, at 1.1 Gyr, a peak is provoked by two different, less massive satellites whose pericenter times are coincident.}

\subsection{Acceleration by gas}
\label{subsec:acceleration_gas}
\begin{figure*}
	\includegraphics[width=1\textwidth]{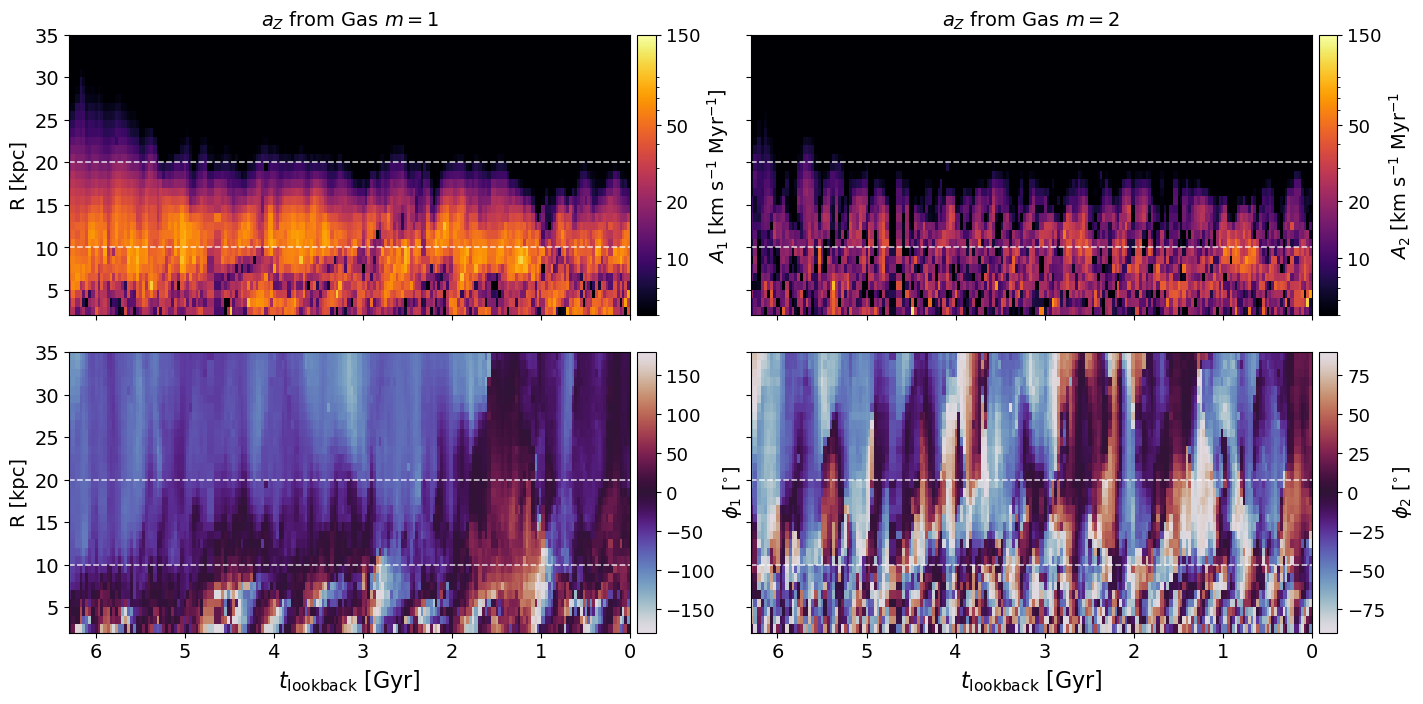}
    \caption{Fourier decomposition of the acceleration of the gas cells (similar to Fig.~\ref{fig:DM_acceleration_fourier}) onto the $Z=0$ plane (galactic plane). We show $m=1$ (left panels) and $m=2$, from which we have the amplitude of the mode (top panels) and their corresponding phase of the maximum amplitude (bottom panels).}
    \label{fig:Gas_acceleration_fourier}
\end{figure*}

We apply the same method as in Sec.~\ref{sec:dm_accelerations} to the gas cells of the simulation, whose results are shown in Fig.~\ref{fig:Gas_acceleration_fourier}. In the bottom left panel, which shows the phase of the mode $m=1$, we see a sharp transition between the outer parts and the inner parts of the disk. This transition is located between 5 and 10 kpc during the first snapshots and between 10 to 15 kpc from 5 Gyr on. This transition can also be seen as a sort of break in the amplitude. 

At the inner regions, between 4.5 and 3.5 Gyr, we observe a retrograde rotating mode $m=1$ that extends to 10 kpc approximately. This pattern has a period between 300 and 500 Myr and is similar to the one of the inner bending wave seen in the left panels of Fig.~\ref{fig:Bending_fourier}. Indeed, we see that at these times and radius, the gaseous disk can be misaligned with the stars by about 4 degrees (top panel of Fig. \ref{fig:Lstars_vs_Lgas_ref_system}), thus causing this specific mode $m=1$ in acceleration.

At the outer parts (at radius of $R>15$ kpc), we observe low amplitudes of the acceleration's mode $m=1$ (top left panel). At these radii, there is a rather constant phase (bottom left panel), which corresponds to an approximate angle of -90$^{\circ}$ (blue colors). In this same panel, from 2 Gyr on, this phase changes to red/black (corresponding to angles between 50$^{\circ}$ and 100$^{\circ}$), propagating towards the outer radius. This is accompanied by a higher amplitude (top left panel). From 1 to 0 Gyr the amplitude of $m=1$ decreases, and the phase shows similar angles to previous instances. To study this in more detail we examined the gas distribution and its temperature in our model. 

%\s{In Fig.~\ref{fig:Gas_edgeon} we show the gaseous disk in edge-on projection (top two rows) and a zoom-out, face-on projection of the temperature of the gas (bottom row panels). In the first column (5.61 Gyr) we see a deformed gaseous disk, tilted from the galactic plane. The zoom-out bottom panel of this snapshot shows the clumps of colder (blue) gas stripped from Arania and the disk itself at its first encounter (at $>$6~Gyr).} 

\begin{figure*}
\centering
	\includegraphics[width=0.88\textwidth]{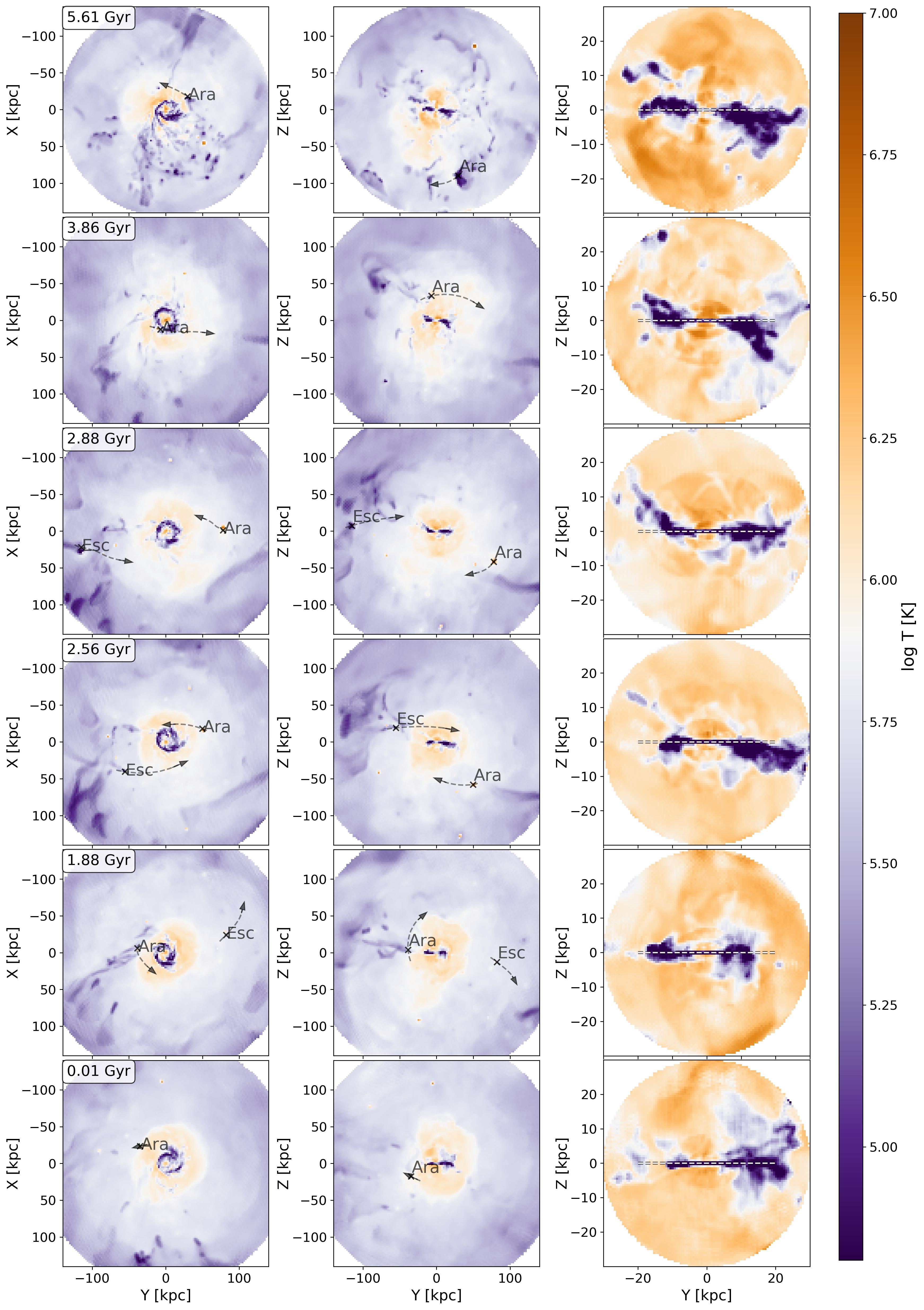}
    \caption{Distribution and temperature of the gas in GARROTXA.
    \textbf{Left and central columns:} Zoom-out face-on ($Y$-$X$ plane) and edge-on ($Y$-$Z$ plane) projection of the gas temperature at six different instances of the simulation colored by temperature. The crosses and dashed lines indicate the position and direction of the satellites Arania and Escarabajo. \textbf{Right column:} Edge-on projection of the gaseous disk, colored by temperature. The dotted horizontal line represents the plane of the stellar disk. The main inflow is reaching the galaxy from the negative Y direction.}
    \label{fig:Gas_edgeon}
\end{figure*}

In Fig.~\ref{fig:Gas_edgeon} we show the gas in a zoom-out, face-on, projection (left column), a zoom-out, edge-on projection (central column), and a zoom-in, edge-on projection of the gaseous disk (right column) colored by temperature. In the first row (5.61 Gyr) we see a highly clumpy distribution of cold (blue) gas, stripped from Arania at its first encounter (at $>$6~Gyr). The right panel of this row (zoom-in) shows a deformed gaseous disk, highly tilted from the galactic plane.

%tilted from the galactic plane. The zoom-out bottom panel of this snapshot shows the clumps of colder (blue) gas stripped from Arania and the disk itself at its first encounter (at $>$6~Gyr).} 
Over time, as seen in the next zoom-out face-on panels rows, the gas falls into the galaxy preferably through a cold (blue) gas filament in the negative $Y$ and positive $X$ direction of -50$^{\circ}$ approximately at radii of $>50$ kpc, and -90$^{\circ}$ approximately at smaller radii throughout almost all snapshots studied. This is especially well seen at the panels of 2.88, 2.56 and 1.88 Gyr (first column of the third, fourth and fifth row respectively, where the gas falls from the left parts of the plane, dominated by blue colors. Indeed, the latter direction of -90$^{\circ}$ is compatible with the observed phase of the acceleration in the Fourier mode $m=1$. This gas is a mixture of the one that has been stripped from satellites (Arania's at early times, mostly tidally in its first pericenter, and ram-pressure stripped from Escarabajo at later times) and the one that outflowed from the disk at an early time through supernovae winds and initially kept in the circumgalactic medium. This metal-rich gas has shorter cooling times and thus falls into the galaxy \citep{dekel2006galaxy}.

In particular, at 2.88 Gyr Escarabajo is already inside the virial radius and, as mentioned above, has suffered ram-pressure stripping due to the presence of a warm-hot gas corona around the galaxy. Gas stripped from Escarabajo mixes with the pre-existing inflowing gas (as seen at 2.88 Gyr and 2.56 Gyr in the third and fourth rows, respectively), cools down and, over the next half gigayear, enters the disk and increases the star formation (see figure 5 in GC22). 
Following this star formation event, from 2 Gyr on, we observe a change in the phase of the $m=1$ mode (in Fig.~\ref{fig:Gas_acceleration_fourier}), along with a higher amplitude of $m=1$ in $a_{Z}$ which propagates from the inner parts towards the outer radii. At these times, the gaseous disk is tilted with respect to the stellar disk due to the recent interaction with the satellite Escarabajo. After examining the snapshots, we observe that there is also an anisotropic ejection of gas from the supernovae winds following the aforementioned star formation event in this direction which might be further perturbing the disk in this direction. As a consequence of these different phenomena, at 1.88 Gyr (right panel of 5th row in Fig. ~\ref{fig:Gas_edgeon}) the gaseous disk is tilted by about 4$^{\circ}$ to 8$^{\circ}$ (as seen in Appendix \ref{sec:misalignment_appendix}, and in Sec. \ref{sec:results}). Along the last gigayear of evolution, star formation is regulated by the galactic fountain leading to new supernovae events (although with weaker mass ejection than in the former case). %\t{These supernovae bubbles can contribute to the acceleration }

%The different phenomena between internal and external radii can cause additional tension on the gaseous disk, leading to the subsequent turbulence in the gaseous disk.

%\begin{figure*}
%	\includegraphics[width=1\textwidth]{Figures/Gas_edgeon_zoom_out.jpg}
%    \caption{Distribution and temperature of gas in GARROTXA. Top and central rows: Edge-on projection of gas at six different instances of the simulation colored by the surface density and the temperature, respectively. The dotted horizontal line represents the plane of the stellar disk. Bottom row: Face-on projection of gas with further zoom-out, colored by temperature.}
 %   \label{fig:Gas_edgeon}
%\end{figure*}

\section{Triggering agents of bending deformation} \label{sec:results}

\begin{figure*}
 \includegraphics[width=1\textwidth]{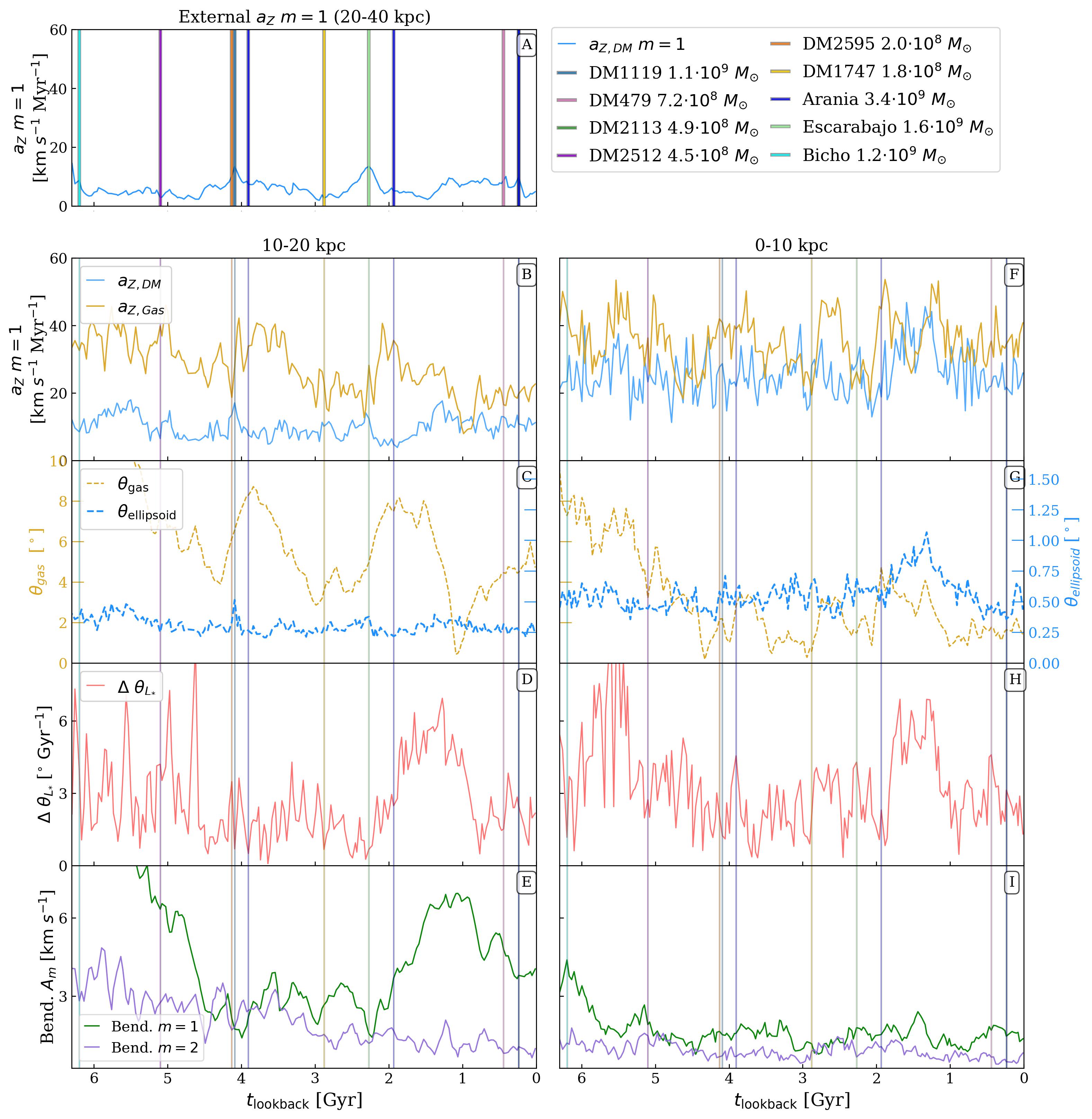}
    \caption{Top left panel (\textbf{A}) mean value of the amplitude of Fourier mode $m=1$ of $a_{Z}$ by the dark matter (blue lines) at external radii (20-40 kpc). The vertical lines show the moments of maximum approach to the disk by those satellites of higher mass, and the labels indicate their $M_{200}$ at infall. The left panels show the calculated properties at radii from 10 to 20 kpc, whereas the right panels show between 0 to 10 kpc. Panels \textbf{B} and \textbf{F} show the mean value of the amplitude of mode $m=1$ of the vertical acceleration of dark matter (blue lines) and gas (orange lines). Panels \textbf{C} and \textbf{G} show the angle difference between the angular momentum of stars and the angular momentum of gas (dotted orange lines, leftmost axis) and the tilt of the central dark matter triaxial structure with respect to the disk (dotted blue lines rightmost axis), denoted by $\theta_{gas}$ and $\theta_{ellipsoid}$, respectively. Panels \textbf{D} and \textbf{H} show the angular momentum vector variation in relation to the previous snapshot divided by the timespan. Panels \textbf{E} and \textbf{I} show the mean amplitude of Fourier amplitudes for bending modes $m=1$ (green lines) and $m=2$ (purple lines).}
    \label{fig:1D_variables}
\end{figure*}

In this section, we make a connection between the times of higher vertical acceleration (from both dark matter and gas) and the behavior of the disk in terms of bending modes calculated in previous sections. In Fig.~\ref{fig:1D_variables} we take the data from Figs. \ref{fig:Bending_fourier}, \ref{fig:DM_acceleration_fourier}, and \ref{fig:Gas_acceleration_fourier}, and get the mean amplitude of the Fourier modes within a radial bin (see top labels) to compare between them at specific radial regions. In panel A we show the mean amplitude of the $m=1$ Fourier mode of the vertical acceleration produced by dark matter, between 20 and 40 kpc. In this region, the effects of satellites and dark matter sub-halos are dominating, as seen in Sec.~\ref{sec:dm_accelerations}, and we see clear peaks at their moments of maximum approach. Panels B and F show the mode $m=1$ of the acceleration by gas (orange line) and dark matter (blue line) at radii from 10 to 20 kpc and 0 to 10 kpc, respectively. We observe that the amplitude of the $m=1$ mode of $a_{Z}$ applied by the gas is as significant as the one by the dark matter and even more dominant in the region from 10 to 20 kpc, due to its strong misalignment with the disk. At radii from 0 to 10 kpc, the $m=1$ acceleration by the gas is comparable to that exerted by the dark matter halo. However, in terms of total vertical acceleration, the inner dark matter triaxial structure dominates due to the contribution of higher modes (Fig. \ref{fig:DM_acceleration_fourier}).

In addition to this, we calculate the angle between the angular momentum vector of the stellar disk and that of the gas measured in spherical shells with radii between 10 to 20 kpc and between 0 and 10 kpc (dashed orange line in panels C and G, respectively). We observe that there is a correlation between the acceleration applied by gas and its angular misalignment with the disk. As mentioned in Sec.~\ref{sec:dm_accelerations}, here we also obtained the tilt of the central dark triaxial structure with respect to the stellar disk by calculating its maximum height and corresponding radius, with the former being the Fourier amplitude of the vertical distribution of dark matter particles in a cylinder with $|Z|< 10$ kpc. Although this tilt is only of 0.5-1.25$^{\circ}$ at times (blue lines in panels C and G),  given the total mass of the triaxial inner dark matter structure this is enough to create large accelerations like the ones observed in Fig.~\ref{fig:DM_acceleration_fourier}. 
We also measure the changes in the angular momentum direction of the stellar disk in relation to the previous snapshot taking the cosmological box as a reference (panels D and H). Panels E and I show the mean amplitudes of the bending modes $m=1$ (green lines) and $m=2$ (purple lines), for regions between 10 and 20 kpc and between 0 and 10 kpc, respectively.

The perturbed initial state of the galaxy and the successive interactions with satellites make it very difficult to pinpoint a moment in which the galaxy is at equilibrium along its history. However, in Sec. \ref{sec:accelerations}, we have described the key moments where interactions with satellites and dark matter sub-halos occur. Based on these, we divide our analysis into two intervals. The first is defined by the interaction with the dark satellites (at 4.1 Gyr), and the other is that of the dwarf satellite galaxy Escarabajo (at 2.2 Gyr). We study both interactions in more detail and their possible consequences on the disk. In Appendix \ref{sec:correlations_appendix} we briefly study how these variables correlate to each other more quantitatively. A more descriptive explanation based on that and Fig. \ref{fig:1D_variables} is presented below.

\subsection{Multiple dark and luminous satellites and bending waves}
\label{subsec:darksatellites}

%\s{At 4.1 Gyr, in Fig.~\ref{fig:Bending_fourier}, we can see that relatively higher amplitudes of bending modes are excited in both the inner ($<$ 10 kpc) and external parts of the disk ($>$ 15 kpc). In the case of outer radii, we observe a retrograde bending mode $m=1$ and prograde bending mode $m=2$, which coincide with the pericenters of two of the dark satellites (DM1119 and DM2595) at 4.1 Gyr. These have, respectively, virial masses $M_{200}$ of $10^9$ and $10^8$ $M_{\odot}$, \t{plane-}crossing positions at a cylindrical galactocentric radius of 20 kpc and 15 kpc, and velocities $V_{Z}$ of $364$ km s$^{-1}$ and $386$ km s$^{-1}$, respectively. Their gravitational acceleration can be seen in the first three top panels in Fig.~\ref{fig:acceleration_example}, and in Fig.~\ref{fig:DM_vs_halos_rockstar}. The pericenter of these satellites are compatible with the increase in bending modes at outer radii. These satellites are also accompanied by the second pericenter of the satellite Arania at 3.9 Gyr (blue line in Fig.~\ref{fig:DM_vs_halos_rockstar}), which now has a virial mass of $M_{200c}$ of $1.1 \times 10^{9}$ $M_{\odot}$, as calculated by Rockstar ($5 \times 10^{8}$ $M_{\odot}$ as calculated in GC22, by measuring the tidal radius at apocenters of the orbit), which may also reinforce the bending patterns.}

At 4.1 Gyr, in Fig.~\ref{fig:Bending_fourier}, we can see that a relatively higher amplitude of bending mode $m=1$ is excited in the inner ($<$ 5 kpc) parts of the disk. In the external parts (> 10 kpc) the increase starts later (from 3.9 Gyr on approximately). Both increases are in relation to previous instances where the amplitude is lower.
. In the case of outer radii, we observe a very slow bending mode $m=1$ and prograde bending mode $m=2$, which coincide with the pericenters of two of the dark satellites (DM1119 and DM2595) at 4.1 Gyr. These have, respectively, virial masses $M_{200}$ of $10^9$ and $10^8$ $M_{\odot}$, plane-crossing positions at cylindrical galactocentric radii of 20 kpc and 15 kpc, and velocities $V_{Z}$ of $364$ km s$^{-1}$ and $386$ km s$^{-1}$, respectively. Their gravitational acceleration can be seen in the first three top panels in Fig.~\ref{fig:acceleration_example}, and in Fig.~\ref{fig:DM_vs_halos_rockstar}. The pericenters of these satellites are compatible with the increase in bending modes at outer radii. During this time, the second pericentric passage of the satellite Arania also takes place at 3.9 Gyr (blue line in Fig.~\ref{fig:DM_vs_halos_rockstar}), which at the time of this interaction has a bound mass $M_{\mathrm{bound}}$ of $1.1 \times 10^{9}$ $M_{\odot}$, as calculated by Rockstar ($5 \times 10^{8}$ $M_{\odot}$ as calculated in GC22, by measuring the tidal radius at apocenters of the orbit), which may also reinforce the bending patterns. 

The inner retrograde bending pattern does not have a clear origin. It seems to be present throughout all the timespan studied. The oscillation of the inner bending mode (panel I of Fig.\ref{fig:1D_variables}) could be related to the inner density structures of the disk (see Appendix \ref{sec:density_breathing_relation}). But, specifically from 4 Gyr, it has a higher amplitude with a maximum at 3.5 Gyr (top left panel of Fig.~\ref{fig:Bending_fourier}) and a more clear phase (bottom left panel of Fig.~\ref{fig:Bending_fourier}). In Sec. \ref{subsec:acceleration_gas} we found that the accelerations from the gas show a wave-like pattern with similar behavior in rotation and periodicity. Its amplitude (panel F) increases approximately 400 Myr before the interaction with the dark satellites DM1119 and DM2595 (blue and orange line respectively, at 4.1 Gyr), and other less massive sub-halos near the disk at that moment. Thus, it is unlikely that these sub-halos are the cause of the increase in inner bending. At these times, we also see a slight misalignment in the gas (panel G). Our hypothesis is that gas is being accreted and destabilizing both the stellar and gaseous disk, inducing these specific bending waves, or at least enhancing the already existing one. 

%\s{ESTO DE AQUÍ YA NO ES CIERTO PARA ESTE SISTEMA DE REFERENCIA. COMO ALINEABAMOS A 15 KPC NO NOS SALÍA EL BENDING DE 10-15, Y A CAMBIO NOS SALÍA LA PARTE INTERNA DESALINEADA, ACOMPAÑADO POR UN DESALINEADO DEL ELIPSOIDE, QUE AHORA YA NO SE VE, LO QUE DA MÁS IMPORTANCIA AL DESALINEADO DEL GAS EN ESTAS REGIONES, QUE SÍ ES REAL EN ESTE SISTEMA DE REFERENCIA TAMBIÉN. At approximately 3.2 Gyr we also see an increase in the global acceleration from the dark matter in the inner parts (solid blue line in panel F) coming from the misalignment with inner ellipsoid (dotted blue line in panel G). This misalignment with the central ellipsoid may reinforce the bending waves at these radii. }

\subsection{Distant interaction triggering bending modes via disk misalignment}
\label{subsec:escarabajo}
The dwarf satellite Escarabajo has a total mass of $M_{200}$ of $1.6\times 10^9$ M$_{\odot}$, and a stellar mass of $6\times 10^6$ M$_{\odot}$ at its first infall, before its closest approach (2.2 Gyr). This interaction is slower ($V_{Z}=-50$ km s$^{-1}$ at the moment of the pericenter) than the ones mentioned in the previous subsection (see Sec.~\ref{subsec:darksatellites}). Escarabajo's pericentric passage occurs at $R=40$ kpc and at 16 kpc above the plane, thus far from it. We saw in Fig.~\ref{fig:DM_vs_halos_rockstar} that the vertical acceleration by this satellite is mostly at the outer disk and barely reaches the inner parts. Even though the amplitude of the bending mode $m=1$ (panel E of Fig.~\ref{fig:1D_variables}) reaches its maximum a Gyr later, its increase starts approximately 300 Myr after the pericenter of Escarabajo, which is half of the rotation period at these outer parts (600 Myr at 20 kpc). Furthermore, we need to take into account that, this being a slower interaction, the acceleration is not only applied at the moment of the pericenter but also several Myr before and after its maximum approach. Thus, this initial increase is compatible with the passing of Escarabajo.

The angular momentum of the stars also starts to change at 300 Myr after Escarabajo’s passing (as seen in panels D and H). This indicates that Escarabajo’s passage pulls and tilts the disk, especially at outer radii, where self-gravity is smaller. It also slightly shifts from its alignment with the central triaxial structure of the dark matter halo, which also generates accelerations that prevail longer in time. This creates a misalignment between the disk and the central dark matter and old stars (panel G) triaxial structure, which also exerts an acceleration into this disk. This is reflected in the amplitude of mode $m=1$ of the acceleration by dark matter (panels B and F, blue lines), which reaches its maximum at approximately 1.2 Gyr.

Furthermore, in Sec. \ref{subsec:acceleration_gas}, we have seen how Escarabajo loses its gas and this mixes with previous existing inflows. We measure the cold gas inside Escarabajo's virial radius before its maximum approach and we find that this satellite contributes to $10^{7}$ $M_{\odot}$ of cold gas ($T<10^{4}$ $K$). At these times, the misalignment of the gas disk in relation to the plane is 8 degrees (panel C of Fig.~\ref{fig:1D_variables}) and has an associated increase in the vertical acceleration by the gas at radii of 10-20 kpc (panel B). This is a period of destabilization of both the gaseous and stellar disk as a consequence of the direct interaction with Escarabajo. Even though the infalling gas may not induce large perturbations by itself due to its mass, this new gas is settling in a different angle in respect to the disk. This new gas forms stars and the subsequent SNe bubbles further perturbs the disk.

The bending mode $m=1$ finally reaches its maximum at approximately 1.1 Gyr. This perturbation is present across stellar populations as a whole, but we find that it is stronger in stellar particles from 4 to 6 Gyr of age, approximately. The phase of this bending mode (bottom left panel of Fig.~\ref{fig:Bending_fourier}) indicates that it is aligned with the major axis of the central triaxial dark matter structure, which at this moment is tilted in the opposite direction. Thus, the central triaxial structure of dark matter halo appears to be crucial to the behavior of the bending pattern once it has been triggered. In Sec.~\ref{sec:discussion}, we also discuss briefly the possible effects of the disk itself on the shape of the bending wave.
%\textbf{Because of the disk’s self-gravity, the inner parts of the disk also exert a torque. The orientation of the resulting spiral pattern (leading or trailing) depends on the relative magnitudes of the two torques \citep{shen2006galactic, gomez2016fully}. In this case, the shape of the warp (leading in this case), indicates that the torque induced by the inner disk dominates after the initial perturbation, which results in the leading shape of the warp.}

Being such a complex system implies the existence of interactions that are amplified (or dampened) by the different components and structures. Thus, the gas and especially the central triaxial dark matter structure seem to contribute to a general destabilization between different parts of the disk, leading to the amplification and prolongation in time of the bending mode. It is expected to have a delay in the large-scale effects caused by the "tug" of satellites if they induce other imbalances. In addition to this, there are also all the new passages of other, less massive, but non-negligible, satellites from 1 Gyr on (see Fig.~\ref{fig:DM_vs_halos_rockstar}).

\section{Discussion and conclusions}\label{sec:discussion}
%First,summarize the key findings from the research and link them to the initial research question. Seek to answer this question: What should readers take away from this paper? 
\subsection{Summary}
The existence of vertical waves and vertical phase mixing in the MW and in other galaxies has motivated our exploration of the bending modes in the stellar disk of the GARROTXA cosmological simulation. Using Fourier decomposition, we analyzed these modes and the vertical gravitational accelerations produced by the dark matter halo, gas, and satellite galaxies. Our main findings on the study of the bending of the stellar disk are:

\begin{enumerate}

   \item The bending behavior of the stellar disk is much more complex than previously seen in N-Body simulations with different regimes at different radii and times. Throughout the simulation, we find different $m=1$ modes coexisting but being exited at different times: 
   One is slower and extends from $\sim$7 kpc to 20 kpc. This external mode is generally aligned with the major axis of the inner structure of old stars and dark matter. We also find an inner, with much lower amplitude, retrograde mode with high frequency dominating at $<$5 kpc.
   %\s{Throughout the simulation we find two inner $m=1$ modes coexisting, but being exited at different times. One retrograde with high frequency dominating at $\sim$5 kpc and another more static and extending to $\sim$10 kpc. There is also an outer (>10 kpc) slower and retrograde mode that most of the time shows a diametrically opposite phase compared to the intermediate disk regions.}
   
   \item The amplitudes of the bending mode are high in the early stages of the thin disk formation (20~km s$^{-1}$) and they can be up to 8.5~km s$^{-1}$ in the late disk evolution.
      
   \item The slow pericentric passage of a satellite galaxy of mass $M_{200}$ of $1.6\times 10^9$ M$_{\odot}$ triggers a disk tilting. When the stellar disk is tilted, it becomes misaligned with the inner dark matter triaxial structure (formed in an ancient major merger) which is less responsive to these perturbations as it is more massive. The effect on the disk bending is magnified by this new misalignment with the dark matter inner structure, leading to the highest vertical deformation in the last several Gyrs of evolution.

  \item When the satellite enters the virial radius of the host, the gas of the satellite is also stripped by ram-pressure and it slowly cools down and falls into the disk along with pre-existing inflows. This cold gas enters the stellar disk non-isotropically, settles in a slightly tilted new gaseous disk, and, lately, induces new star formation events and the creation of supernovae bubbles. Even though these processes may have a weaker impact on the disk, they are compatible with the accelerations seen exerted by the gaseous disk. Thus, all these processes might magnify the initial bend.
  
 % \s{\item Another satellite galaxy is ram-pressure stripped at large radii and its gas slowly falls in along with pre-existing inflows. This cold gas enters the stellar disk non-isotropically, settles in a slightly tilted new gaseous disk, and induces new star formation events and the creation of supernovae bubbles afterward. All these processes have an impact on the original stellar disk structure which bends.}

 %PROPUESTA \t {

   \item After one of the several interactions with dark sub-halos, the inner $m=1$ retrograde bending mode increases its amplitude and shows a phase pattern that correlates with the one of the gas vertical acceleration. Although this amplitude is relatively lower than the external bending discussed above, this may suggest that the gas-stellar disk misalignment plays an important role in the observed inner disk bending.
    %\item When the stellar disk tilts, it becomes misaligned with the inner stellar/dark matter triaxial structure formed in a previous major merger, which is much less responsive to perturbations due to its large mass. The previously mentioned effect on the disk structure is magnified by this additional misalignment that occurs in the disk with respect to the central dark matter structure. This also applies a vertical acceleration onto the disk, leading to the highest vertical deformation in the last several Gyrs of evolution.

\end{enumerate}

%Second,place the findings in context. This step will involve going back to the literature review section and analyzing how the results fit in with previous research. 
%Third,mention and discuss any unexpected results.Describe the results and provide a reasonable interpretation of why they may have appeared. Additionally, if an unexpected result is significant to the research question, be sure to explain that connection. Discussion Section for Research Papers, Fall 2021.1of 5

\subsection{Discussion}
We thus identified the following agents that trigger the bending of the stellar disk: 

\begin{enumerate}
    \item \emph{Dark sub-halos}: Dark satellites with masses $\sim10^9$  $M_{\odot}$ interact with the stellar disk on many occasions coinciding with a higher amplitude of the bending mode $m=1$. The presence of such dark sub-halos is expected in the $\Lambda$CDM galaxy formation theories and their interaction with the disk has been mostly studied as a source of disk heating (see e.g. \citealt{benson2004heating, d2016excitation}). Here we highlight the role of such structures as triggers of bending perturbations of the disk.

    \item \emph{Satellite galaxies}: Satellite galaxies in the GARROTXA simulation at z$>2$ have at most M$_{200}\sim$10$^{10}$ M$_{\odot}$ before their first infall, and most of them do not reach inner regions of the galaxy. However, although their direct tidal influence in the inner stellar disk is not large, they lead to a cascade of events that trigger some perturbations as discussed below.

  %  \item \emph{Inner dark matter profile:} The inner tri-axial structure of the dark matter is an imprint of its ancient accretion history. At certain times, the largest vertical gravitational acceleration comes from a misalignment between the stellar disk and this dark matter ellipsoid. The misalignment is triggered by the tilting of the stellar disk, and the resulting bending can last for more than 1 Gyr. Similar phenomena have also been observed in \citep{sellwood2022internally}, where slow bending waves arise from misalignment between the inner and outer regions of the disk. We cannot discard an intrinsic coupling mode between the disk and halo contributing to this deformation. 

    \item \emph{Inner density profile:} The inner tri-axial structure of the dark matter is an imprint of its ancient accretion history. At certain times, the largest vertical gravitational acceleration comes from a misalignment between the stellar disk and this dark matter ellipsoid. The misalignment is triggered by the tilting of the stellar disk, and the resulting bending can last for more than 1 Gyr. Similar phenomena have also been observed in \citet{sellwood2022internally}, where slow bending waves arise from misalignment between the inner and outer regions of the disk. We cannot discard an intrinsic coupling mode between the disk and halo contributing to this deformation. Furthermore, it is possible that the overall inner density profile of the disk also exerts a torque after the initial disk tilting, resulting in the leading shape of the warp \citep{shen2006galactic, gomez2016fully}, which is worth exploring in future analysis.
        
    \item \emph{Gas accretion and misaligned gaseous structures:} The vertical accelerations from the gaseous structures, often misaligned with the stellar disk, are comparable to those from satellites and dark matter. The origin of this misalignment is the inhomogeneity of the cold flows, the slightly different angular momentum of the incoming gas with respect to the actual stellar disk, and the expansion of supernovae bubbles. We speculate that all these phenomena may have a further impact on the bending behavior of the stellar disk, although we cannot measure in which exact proportion. The effect of extra-planar non-homogeneous distributed gas has also been detected as a source of vertical patterns and disk tilting in other models \citep[e.g.][]{gomez2017,khachaturyants2022bending}.
\end{enumerate}

Below we discuss other perturbation mechanisms that we have not looked into in detail but may need special consideration:

\begin{enumerate}
\setcounter{enumi}{4}
    \item \emph {Wake:} We have not found a dominant global asymmetry like the one expected when a strong wake is present in the Fourier analysis of the GARROTXA model. This could be because there is only one satellite galaxy with comparable mass to the Sagittarius Dwarf Galaxy, and its pericenter occurs at large radii from the main host's center. This satellite has a mass above $10^{10}$ $M_{\odot}$ before entering the virial radius of the host for the first time, and it loses mass down to 4.7$\times10^9$~$M_{\odot}$ at the moment of the first pericenter. Since the orbit of the satellite is mostly polar and slightly retrograde, we expect most of the torques to be in the vertical direction. However, in our case, both the collective and transient response to the wake would be weaker than expected for a Sagittarius-like satellite. Even if we have not quantified the exact torque applied by the wake, its effects are already captured by our calculation of accelerations by all dark matter particles. Furthermore, due to the dipolar nature of the halo's response, the wake effects are accounted for in our Fourier analysis. 

%debris
    \item \emph {Streams:} We also have studied the tidal streams from Arania after its first pericenter (both stellar and dark matter debris) and we have found that the maximum vertical acceleration applied by this material is negligible in comparison with the progenitor and the rest of the dark matter halo. However, in the case of more massive satellites, their debris could be significant enough to affect the general structure of the dark matter halo \citep{garavito2021quantifying}.
%... Even though is not as high as the acceleration applied by a satellite at the moment of its pericenter, it applies a constant acceleration and it can apply XX\% of the satellite's vertical acceleration at its maximum. 

%density disk structures
    \item \emph {Density structures of the disk:} Other structures present in the GARROTXA model are a central stellar oval and a one-armed retrograde density pattern (see Appendix~\ref{sec:density_breathing_relation}). We have found some weak correlations between these stellar disk's non-axisymmetric structures and the breathing mode. Thus, we suspect that the physics of the vertical behavior in our model is also related to the density structure of the disc. However, we will study how these self-gravitating structures can perturb the disk in the future (see e.g. \citealt{grion2021holistic} for an analysis in a pure N-Body simulation).

\end{enumerate}

In GC22 we detected an increase in the amplitude of the phase spirals in intermediate-age stars coinciding with the most important satellites' percienters. We observe that the periodicity of the thick bands present in Figure 5 of GC22 his of approximately 1.2 Gyr, which is roughly coincident with the period of the one-armed density mode (Appendix ~\ref{sec:density_breathing_relation}). This suggests that the phase spiral may propagate as this density mode in our model. The observed phase spirals are related to the phase mixing of the bending modes of the disc studied here, even if these appear to be not very large compared to other models (amplitudes of 8.5 km s$^{-1}$ in this model compared to amplitudes up to 30 km s$^{-1}$ in \citealt{gomez2017}). In \cite{tremaine2023origin} they propose a mixed scenario with several large-scale perturbations and numerous small-scale perturbations. This is qualitatively compatible with our model in the sense of having multiple dark-(and not dark) satellites affecting the disk. However, due to the resolution limitations, and the several interactions occurring simultaneously, we could not accurately track the spiral pattern or measure the winding of the phase spiral and, except for some temporal coincidences, we cannot attribute them to a specific event. The conclusions of this work point also to a mix of many processes (not present in the model of \citealt{tremaine2023origin}) such as the structure of dark matter halo, accretion history, gas content, etc.

We conclude that although further study is needed, the phase spirals we observed can be a consequence of the many perturbations that the stellar disk suffers directly or indirectly related to the satellite galaxies' infall but also gas accelerations and inner dark matter structure.

%caveats
\subsection{Caveats}

There are also some caveats in our study that need to be considered.  
For our analysis, we define a new reference system at each timestep. However, it is not trivial to define a midplane with such a complex vertical structure, with different bending at different radii or for different populations. Variations in the determination of the galactic plane can lead to inaccuracies in the measurements of dynamical properties \citep{beane2019implications}.

The application of the tools used in this work is necessary to detect and analyze the variety of processes taking place in cosmological simulations. They are also necessary in order to analyze a larger set of models, which is crucial for a more global understanding of disk dynamics in real galaxies. However, another caveat in our methodology is that our Fourier analysis does not always provide conclusive evidence to establish true causality. This likely comes from two aspects. Firstly, when there are many processes in place, and some of them are complexly interconnected, it is not straightforward to distinguish between real agents of perturbation or simple coexisting effects, or even to separate causes and their effects. Secondly, this limitation might also be related to the Fourier analysis. Localized accelerations like those from sub-halos and satellite galaxies will be somehow smoothed by the sine/cosine decomposition, especially when we only keep the series to the first orders ($m=1$ and $m=2$ modes). In the case of the internal non-axisymmetric dark matter structure, its total acceleration is not completely captured by the $m=1$ mode, since it is also dominant in higher modes. Finally, since there are different disk regimes with radial limits that change with time, our analysis can be biased by exploring the disk in rings of fixed radius that might not be the natural ones. 

To overcome these limitations, in the future, it might be worth exploring Basis Function Expansions or Multichannel Singular Spectrum Analysis for both the acceleration field and the disk kinematic perturbations. More sophisticated tools to link the different agents and effects might improve our correlation analysis (Appendix \ref{sec:correlations_appendix}). These techniques are starting to be applied to disc dynamics in more controlled simulations \citep{weinberg2021using,garavito2021quantifying, johnson2023dynamical} and this might establish the foundations for its application in more complex cases such as the model used here. These expansions might be also useful for a more global study of the dynamics, not centered exclusively on the vertical or planar dimensions. %Other techniques such as Dynamic Mode Decomposition \citep{darling2019eigenfunctions} might be useful in such complex analysis.

Also, even if the number of satellites in this model is compatible with the predictions for the Milky Way's mass \citep{bullock2017small}, a lack of interaction with massive satellites (M$_{*}$>$10^{10}$  $M_{\odot}$) in this system that could be equivalent to the LMC and SMC makes it challenging to compare directly with the Milky Way at the present time. However, this exercise helps us to understand the evolution of disk galaxies in a global way, where past interactions can have a long-lasting impact on stellar disk morphology and kinematics.

\subsection{Final conclusions and remarks}

%Fifth, provide a brief look at potential follow-up research studies. Recommend a few areas where further investigation may be crucial. However, don’t go overboard with the suggestions, as they can leave a reader thinking more about the gaps in the paper rather than the actual findings. 

%Sixth (and finally),conclude with are statement of the most significant findings and their implications. Explain why the research is important and remind readers of the connections it has to outside material, such as existing literature or an aspect of the field that is affected by the study.
%Synthesize. don't Summarize
%Add significance. don't Repeat yourself
%Reiterate the importance of your message. don't Add new information
%Leave your reader with an impression

A general conclusion of our work is that even galaxies that are far from dense groups can be highly and complexly perturbed systems in a realistic cosmological context, never reaching dynamical equilibrium. We adopt the idea by \cite{grion2021holistic} of the importance of a \emph{holistic} approach to disc dynamics, to which we add here additional factors: that of the effects of the inner profiles of dark matter and old stars from previous major mergers, and the gas brought in by satellites and accreted through filaments. These factors are often overlooked by more simplistic models that therefore do not capture the full complexity of galactic systems.

It is undoubtedly a huge effort to bring cosmological simulations into the picture of disk dynamics. Although detecting and understanding the complex interplay between all existing phenomena and the different dynamical regimes in the disc in cosmological simulations is challenging, the analysis of coexisting perturbative processes will lead to more realistic galactoseismology.

\begin{acknowledgements}
We thank Marcel Bernet, Merce Romero-Gomez, Oscar Jimenez, and Francesca Fragkoudi for discussions about the reference system. We also thank Esperanza Mur for her initial studies of the Fourier density modes of this model. BGC and SRF work has been supported by the Madrid Government (Comunidad de Madrid-Spain) under the Multiannual Agreement with Complutense University in the line Program to Stimulate Research for Young Doctors in the context of the V PRICIT. They also acknowledge financial support from the Spanish Ministry of Economy and Competitiveness (MINECO) under grant number AYA2016-75808-R, AYA2017-90589-REDT, RTI2018-096188-B-I00 and S2018/NMT-429, and from the CAM-UCM under grant number PR65/19-22462. BGC acknowledges IPARCOS Institute for the grant "Ayudas de doctorado IPARCOS-UCM/2022". TA acknowledges the grant RYC2018-025968-I funded by MCIN/AEI/10.13039/501100011033 and by ``ESF Investing in your future''. This work was (partially) supported by the Spanish MICIN/AEI/10.13039/501100011033 and by "ERDF A way of making Europe" by the “European Union” and the European Union «Next Generation EU»/PRTR, through grants PID2021-125451NA-I00 and CNS2022-135232, and the Institute of Cosmos Sciences University of Barcelona (ICCUB, Unidad de Excelencia ’Mar\'{\i}a de Maeztu’) through grant CEX2019-000918-M.
SRF also acknowledges support from a Spanish postdoctoral fellowship, under grant number 2017-T2/TIC-5592.
SRF acknowledges support from the Knut and Alice Wallenberg Foundation and the Swedish Research Council (grant 2019-04659).
PR acknowledges support by the Agence Nationale de la Recherche (ANR project SEGAL ANR-19-CE31-0017 and project ANR-18-CE31-0006) as well as from the European Research Council (ERC grant agreement No. 834148). Simulations were performed on the {\sc Miztli} supercomputer at the LANACAD, UNAM, within the research project LANCAD-UNAM-DGTIC-151. FAG acknowledges support from ANID FONDECYT Regular 1211370, the Max Planck Society through a “Partner Group” grant and ANID Basal Project FB210003. MAGF and BGC acknowledge financial support from the Spanish Ministry of Science and Innovation through the project PID2020-114581GB-C22. \\

\textit{Software: } \textit{yt}  \citep{yt},  \textsc{numpy}\citep{van2011numpy},
\textsc{scipy} \citep{jones2001scipy}, 
\textsc{scikit-learn} \citep{pedregosa2011scikit, buitinck2013api}. 
This research has made use of NASAs Astrophysics Data System (ADS), and the curated research-sharing platform arXiv.
\end{acknowledgements}

\section*{Data availability}

The data underlying this article will be shared on reasonable request to the corresponding author.

\bibliographystyle{aa} % style aa.bst
\bibliography{main} % your references Yourfile.bib

\begin{thebibliography}{66}
\expandafter\ifx\csname natexlab\endcsname\relax\def\natexlab#1{#1}\fi

\bibitem[{Alinder {et~al.}(2023)Alinder, McMillan, \&
  Bensby}]{alinder2023investigating}
Alinder, S., McMillan, P., \& Bensby, T. 2023, arXiv preprint arXiv:2303.18040

\bibitem[{{Antoja} {et~al.}(2018){Antoja}, {Helmi}, {Romero-G{\'o}mez}, {Katz},
  {Babusiaux}, {Drimmel}, {Evans}, {Figueras}, {Poggio}, {Reyl{\'e}}, {Robin},
  {Seabroke}, \& {Soubiran}}]{antoja18}
{Antoja}, T., {Helmi}, A., {Romero-G{\'o}mez}, M., {et~al.} 2018, \nat, 561,
  360

\bibitem[{Antoja {et~al.}(2022)Antoja, Ramos, Garc{\'\i}a-Conde, Bernet,
  Laporte, \& Katz}]{antoja2022phase}
Antoja, T., Ramos, P., Garc{\'\i}a-Conde, B., {et~al.} 2022, arXiv preprint
  arXiv:2212.11987

\bibitem[{Beane {et~al.}(2019)Beane, Sanderson, Ness, Johnston, Grion~Filho,
  Mac~Low, Angl{\'e}s-Alc{\'a}zar, Hogg, \& Laporte}]{beane2019implications}
Beane, A., Sanderson, R.~E., Ness, M.~K., {et~al.} 2019, The Astrophysical
  Journal, 883, 103

\bibitem[{Behroozi {et~al.}(2012)Behroozi, Wechsler, \&
  Wu}]{behroozi2012rockstar}
Behroozi, P.~S., Wechsler, R.~H., \& Wu, H.-Y. 2012, The Astrophysical Journal,
  762, 109

\bibitem[{{Bennett} \& {Bovy}(2021)}]{Bennett2021}
{Bennett}, M. \& {Bovy}, J. 2021, \mnras, 503, 376

\bibitem[{Benson {et~al.}(2004)Benson, Lacey, Frenk, Baugh, \&
  Cole}]{benson2004heating}
Benson, A., Lacey, C., Frenk, C., Baugh, C., \& Cole, S. 2004, Monthly Notices
  of the Royal Astronomical Society, 351, 1215

\bibitem[{{Binney} \& {Sch{\"o}nrich}(2018)}]{binney2018origin}
{Binney}, J. \& {Sch{\"o}nrich}, R. 2018, \mnras, 481, 1501

\bibitem[{{Bland-Hawthorn} {et~al.}(2019){Bland-Hawthorn}, {Sharma},
  {Tepper-Garcia}, {Binney}, {Freeman}, {Hayden}, {Kos}, {De Silva}, {Ellis},
  {Lewis}, {Asplund}, {Buder}, {Casey}, {D'Orazi}, {Duong}, {Khanna}, {Lin},
  {Lind}, {Martell}, {Ness}, {Simpson}, {Zucker}, {Zwitter}, {Kafle},
  {Quillen}, {Ting}, \& {Wyse}}]{bland2019galah}
{Bland-Hawthorn}, J., {Sharma}, S., {Tepper-Garcia}, T., {et~al.} 2019, \mnras,
  486, 1167

\bibitem[{{Bland-Hawthorn} \& {Tepper-Garc{\'\i}a}(2021)}]{Bland-Hawthorn2021}
{Bland-Hawthorn}, J. \& {Tepper-Garc{\'\i}a}, T. 2021, \mnras
  [\eprint[arXiv]{2009.02434}]

\bibitem[{Brown {et~al.}(2021)Brown, Vallenari, Prusti, De~Bruijne, Babusiaux,
  Biermann, Creevey, Evans, Eyer, Hutton, {et~al.}}]{collaboration2021gaia}
Brown, A.~G., Vallenari, A., Prusti, T., {et~al.} 2021, Astronomy \&
  Astrophysics, 649, A1

\bibitem[{{Bryan} \& {Norman}(1998)}]{bryan1998statistical}
{Bryan}, G.~L. \& {Norman}, M.~L. 1998, \apj, 495, 80

\bibitem[{Buitinck {et~al.}(2013)Buitinck, Louppe, Blondel, Pedregosa, Mueller,
  Grisel, Niculae, Prettenhofer, Gramfort, Grobler, {et~al.}}]{buitinck2013api}
Buitinck, L., Louppe, G., Blondel, M., {et~al.} 2013, arXiv preprint
  arXiv:1309.0238

\bibitem[{Bullock \& Boylan-Kolchin(2017)}]{bullock2017small}
Bullock, J.~S. \& Boylan-Kolchin, M. 2017, Annual Review of Astronomy and
  Astrophysics, 55, 343

\bibitem[{Carrillo {et~al.}(2018)Carrillo, Minchev, Kordopatis, Steinmetz,
  Binney, Anders, Bienayme, Bland-Hawthorn, Famaey, Freeman,
  {et~al.}}]{carrillo2018milky}
Carrillo, I., Minchev, I., Kordopatis, G., {et~al.} 2018, Monthly Notices of
  the Royal Astronomical Society, 475, 2679

\bibitem[{{Chequers} {et~al.}(2018){Chequers}, {Widrow}, \&
  {Darling}}]{Chequers2018}
{Chequers}, M.~H., {Widrow}, L.~M., \& {Darling}, K. 2018, \mnras, 480, 4244

\bibitem[{{Darling} \& {Widrow}(2019)}]{Darling2019}
{Darling}, K. \& {Widrow}, L.~M. 2019, \mnras, 484, 1050

\bibitem[{de~la Vega {et~al.}(2015)de~la Vega, Quillen, Carlin, Chakrabarti, \&
  D'Onghia}]{de2015phase}
de~la Vega, A., Quillen, A.~C., Carlin, J.~L., Chakrabarti, S., \& D'Onghia, E.
  2015, Monthly Notices of the Royal Astronomical Society, 454, 933

\bibitem[{Debattista(2014)}]{debattista2014vertical}
Debattista, V.~P. 2014, Monthly Notices of the Royal Astronomical Society:
  Letters, 443, L1

\bibitem[{Dekel \& Birnboim(2006)}]{dekel2006galaxy}
Dekel, A. \& Birnboim, Y. 2006, Monthly notices of the royal astronomical
  society, 368, 2

\bibitem[{D’Onghia {et~al.}(2016)D’Onghia, Madau, Vera-Ciro, Quillen, \&
  Hernquist}]{d2016excitation}
D’Onghia, E., Madau, P., Vera-Ciro, C., Quillen, A., \& Hernquist, L. 2016,
  The Astrophysical Journal, 823, 4

\bibitem[{Faure {et~al.}(2014)Faure, Siebert, \& Famaey}]{faure2014radial}
Faure, C., Siebert, A., \& Famaey, B. 2014, Monthly Notices of the Royal
  Astronomical Society, 440, 2564

\bibitem[{Feldmann \& Spolyar(2015)}]{feldmann2015detecting}
Feldmann, R. \& Spolyar, D. 2015, Monthly Notices of the Royal Astronomical
  Society, 446, 1000

\bibitem[{Frankel {et~al.}(2023)Frankel, Bovy, Tremaine, \&
  Hogg}]{frankel2023vertical}
Frankel, N., Bovy, J., Tremaine, S., \& Hogg, D.~W. 2023, Monthly Notices of
  the Royal Astronomical Society, 521, 5917

\bibitem[{{Gaia Collaboration} {et~al.}(2018){Gaia Collaboration}, {Brown},
  {Vallenari}, {Prusti}, {de Bruijne}, {Babusiaux}, {Bailer-Jones}, {Biermann},
  {Evans}, {Eyer}, \& et~al.}]{gaia2018gaia}
{Gaia Collaboration}, {Brown}, A.~G.~A., {Vallenari}, A., {et~al.} 2018, \aap,
  616, A1

\bibitem[{{Gandhi} {et~al.}(2021){Gandhi}, {Johnston}, {Hunt}, {Price-Whelan},
  {Laporte}, \& {Hogg}}]{gandhi2021snails}
{Gandhi}, S.~S., {Johnston}, K.~V., {Hunt}, J. A.~S., {et~al.} 2021, arXiv
  e-prints, arXiv:2107.03562

\bibitem[{Garavito-Camargo {et~al.}(2021)Garavito-Camargo, Besla, Laporte,
  Price-Whelan, Cunningham, Johnston, Weinberg, \&
  G{\'o}mez}]{garavito2021quantifying}
Garavito-Camargo, N., Besla, G., Laporte, C.~F., {et~al.} 2021, The
  Astrophysical Journal, 919, 109

\bibitem[{Garc{\'\i}a-Conde {et~al.}(2022)Garc{\'\i}a-Conde, Roca-F{\`a}brega,
  Antoja, Ramos, \& Valenzuela}]{garcia2022phase}
Garc{\'\i}a-Conde, B., Roca-F{\`a}brega, S., Antoja, T., Ramos, P., \&
  Valenzuela, O. 2022, Monthly Notices of the Royal Astronomical Society, 510,
  154

\bibitem[{{G{\'o}mez} {et~al.}(2013){G{\'o}mez}, {Minchev}, {O'Shea}, {Beers},
  {Bullock}, \& {Purcell}}]{Gomez2013}
{G{\'o}mez}, F.~A., {Minchev}, I., {O'Shea}, B.~W., {et~al.} 2013, \mnras, 429,
  159

\bibitem[{{G{\'o}mez} {et~al.}(2021){G{\'o}mez}, {Torres-Flores},
  {Mora-Urrejola}, {Monachesi}, {White}, {Maffione}, {Grand}, {Marinacci},
  {Pakmor}, {Springel}, {Frenk}, {Amram}, {Epinat}, \& {Mendes de
  Oliveira}}]{Gomez2021}
{G{\'o}mez}, F.~A., {Torres-Flores}, S., {Mora-Urrejola}, C., {et~al.} 2021,
  \apj, 908, 27

\bibitem[{G{\'o}mez {et~al.}(2016)G{\'o}mez, White, Marinacci, Slater, Grand,
  Springel, \& Pakmor}]{gomez2016fully}
G{\'o}mez, F.~A., White, S.~D., Marinacci, F., {et~al.} 2016, Monthly Notices
  of the Royal Astronomical Society, 456, 2779

\bibitem[{{G{\'o}mez} {et~al.}(2017){G{\'o}mez}, {White}, {Grand}, {Marinacci},
  {Springel}, \& {Pakmor}}]{gomez2017}
{G{\'o}mez}, F.~A., {White}, S. D.~M., {Grand}, R. J.~J., {et~al.} 2017,
  \mnras, 465, 3446

\bibitem[{Grand {et~al.}(2022)Grand, Pakmor, Fragkoudi, G{\'o}mez, Trick,
  Simpson, van~de Voort, \& Bieri}]{grand2022dark}
Grand, R.~J., Pakmor, R., Fragkoudi, F., {et~al.} 2022, arXiv preprint
  arXiv:2211.08437

\bibitem[{Grion~Filho {et~al.}(2021)Grion~Filho, Johnston, Poggio, Laporte,
  Drimmel, \& D’Onghia}]{grion2021holistic}
Grion~Filho, D., Johnston, K.~V., Poggio, E., {et~al.} 2021, Monthly Notices of
  the Royal Astronomical Society, 507, 2825

\bibitem[{Hunt {et~al.}(2022)Hunt, Price-Whelan, Johnston, \&
  Darragh-Ford}]{hunt2022multiple}
Hunt, J.~A., Price-Whelan, A.~M., Johnston, K.~V., \& Darragh-Ford, E. 2022,
  Monthly Notices of the Royal Astronomical Society: Letters, 516, L7

\bibitem[{{Hunt} {et~al.}(2021){Hunt}, {Stelea}, {Johnston}, {Gandhi},
  {Laporte}, \& {B{\'e}dorf}}]{hunt2021resolving}
{Hunt}, J. A.~S., {Stelea}, I.~A., {Johnston}, K.~V., {et~al.} 2021, \mnras,
  508, 1459

\bibitem[{Johnson {et~al.}(2023)Johnson, Petersen, Johnston, \&
  Weinberg}]{johnson2023dynamical}
Johnson, A., Petersen, M.~S., Johnston, K.~V., \& Weinberg, M.~D. 2023, arXiv
  preprint arXiv:2301.02256

\bibitem[{Jones {et~al.}(2001)Jones, Oliphant, Peterson,
  {et~al.}}]{jones2001scipy}
Jones, E., Oliphant, T., Peterson, P., {et~al.} 2001

\bibitem[{Khachaturyants {et~al.}(2022)Khachaturyants, Beraldo~e Silva,
  Debattista, \& Daniel}]{khachaturyants2022bending}
Khachaturyants, T., Beraldo~e Silva, L., Debattista, V.~P., \& Daniel, K.~J.
  2022, Monthly Notices of the Royal Astronomical Society, 512, 3500

\bibitem[{Kumar {et~al.}(2022)Kumar, Ghosh, Kataria, Das, \&
  Debattista}]{kumar2022excitation}
Kumar, A., Ghosh, S., Kataria, S.~K., Das, M., \& Debattista, V.~P. 2022,
  Monthly Notices of the Royal Astronomical Society, 516, 1114

\bibitem[{Laporte {et~al.}(2018)Laporte, Johnston, G{\'o}mez, Garavito-Camargo,
  \& Besla}]{laporte2018b}
Laporte, C.~F., Johnston, K.~V., G{\'o}mez, F.~A., Garavito-Camargo, N., \&
  Besla, G. 2018, Monthly Notices of the Royal Astronomical Society, 481, 286

\bibitem[{{Laporte} {et~al.}(2018){Laporte}, {Johnston}, {G{\'o}mez},
  {Garavito-Camargo}, \& {Besla}}]{Laporte2018}
{Laporte}, C. F.~P., {Johnston}, K.~V., {G{\'o}mez}, F.~A., {Garavito-Camargo},
  N., \& {Besla}, G. 2018, \mnras, 481, 286

\bibitem[{{Laporte} {et~al.}(2019){Laporte}, {Minchev}, {Johnston}, \&
  {G{\'o}mez}}]{laporte2019footprints}
{Laporte}, C. F.~P., {Minchev}, I., {Johnston}, K.~V., \& {G{\'o}mez}, F.~A.
  2019, \mnras, 485, 3134

\bibitem[{{Li} \& {Shen}(2020)}]{Li2020}
{Li}, Z.-Y. \& {Shen}, J. 2020, \apj, 890, 85

\bibitem[{Monari {et~al.}(2015)Monari, Famaey, \& Siebert}]{monari2015vertical}
Monari, G., Famaey, B., \& Siebert, A. 2015, Monthly Notices of the Royal
  Astronomical Society, 452, 747

\bibitem[{Nandakumar {et~al.}(2022)Nandakumar, Narayan, \&
  Dutta}]{nandakumar2022bending}
Nandakumar, M., Narayan, C., \& Dutta, P. 2022, Monthly Notices of the Royal
  Astronomical Society, 513, 3065

\bibitem[{Pedregosa {et~al.}(2011)Pedregosa, Varoquaux, Gramfort, Michel,
  Thirion, Grisel, Blondel, Prettenhofer, Weiss, Dubourg,
  {et~al.}}]{pedregosa2011scikit}
Pedregosa, F., Varoquaux, G., Gramfort, A., {et~al.} 2011, the Journal of
  machine Learning research, 12, 2825

\bibitem[{Power {et~al.}(2003)Power, Navarro, Jenkins, Frenk, White, Springel,
  Stadel, \& Quinn}]{power2003inner}
Power, C., Navarro, J.~F., Jenkins, A., {et~al.} 2003, Monthly Notices of the
  Royal Astronomical Society, 338, 14

\bibitem[{{Roca-F{\`a}brega} {et~al.}(2016){Roca-F{\`a}brega}, {Valenzuela},
  {Col{\'\i}n}, {Figueras}, {Krongold}, {Vel{\'a}zquez}, {Avila-Reese}, \&
  {Ibarra-Medel}}]{garrotxa}
{Roca-F{\`a}brega}, S., {Valenzuela}, O., {Col{\'\i}n}, P., {et~al.} 2016,
  \apj, 824, 94

\bibitem[{Rocha {et~al.}(2012)Rocha, Peter, \& Bullock}]{rocha2012infall}
Rocha, M., Peter, A.~H., \& Bullock, J. 2012, Monthly Notices of the Royal
  Astronomical Society, 425, 231

\bibitem[{Sch{\"o}nrich \& Dehnen(2018)}]{schonrich2018warp}
Sch{\"o}nrich, R. \& Dehnen, W. 2018, Monthly Notices of the Royal Astronomical
  Society, 478, 3809

\bibitem[{Sellwood \& Debattista(2022)}]{sellwood2022internally}
Sellwood, J. \& Debattista, V.~P. 2022, Monthly Notices of the Royal
  Astronomical Society, 510, 1375

\bibitem[{Semczuk {et~al.}(2020)Semczuk, {\L}okas, D’Onghia, Athanassoula,
  Debattista, \& Hernquist}]{semczuk2020tidally}
Semczuk, M., {\L}okas, E.~L., D’Onghia, E., {et~al.} 2020, Monthly Notices of
  the Royal Astronomical Society, 498, 3535

\bibitem[{Shen \& Sellwood(2006)}]{shen2006galactic}
Shen, J. \& Sellwood, J. 2006, Monthly Notices of the Royal Astronomical
  Society, 370, 2

\bibitem[{Trelles {et~al.}(2022)Trelles, Valenzuela, Roca-F{\'a}brega, \&
  Vel{\'a}zquez}]{trelles2022concurrent}
Trelles, A., Valenzuela, O., Roca-F{\'a}brega, S., \& Vel{\'a}zquez, H. 2022,
  Astronomy \& Astrophysics, 668, A20

\bibitem[{Tremaine {et~al.}(2023)Tremaine, Frankel, \&
  Bovy}]{tremaine2023origin}
Tremaine, S., Frankel, N., \& Bovy, J. 2023, Monthly Notices of the Royal
  Astronomical Society, 521, 114

\bibitem[{{Turk} {et~al.}(2011){Turk}, {Smith}, {Oishi}, {Skory}, {Skillman},
  {Abel}, \& {Norman}}]{yt}
{Turk}, M.~J., {Smith}, B.~D., {Oishi}, J.~S., {et~al.} 2011, The Astrophysical
  Journal Supplement Series, 192, 9

\bibitem[{Urrejola-Mora {et~al.}(2022)Urrejola-Mora, G{\'o}mez, Torres-Flores,
  Amram, Epinat, Monachesi, Marinacci, \& de~Oliveira}]{urrejola2022winds}
Urrejola-Mora, C., G{\'o}mez, F.~A., Torres-Flores, S., {et~al.} 2022, The
  Astrophysical Journal, 935, 20

\bibitem[{van~de Voort {et~al.}(2015)van~de Voort, Davis, Kere{\v{s}},
  Quataert, Faucher-Giguere, \& Hopkins}]{van2015creation}
van~de Voort, F., Davis, T.~A., Kere{\v{s}}, D., {et~al.} 2015, Monthly Notices
  of the Royal Astronomical Society, 451, 3269

\bibitem[{Van Der~Walt {et~al.}(2011)Van Der~Walt, Colbert, \&
  Varoquaux}]{van2011numpy}
Van Der~Walt, S., Colbert, S.~C., \& Varoquaux, G. 2011, Computing in science
  \& engineering, 13, 22

\bibitem[{{Weinberg}(1998)}]{Weinberg1998}
{Weinberg}, M.~D. 1998, \mnras, 299, 499

\bibitem[{Weinberg \& Petersen(2021)}]{weinberg2021using}
Weinberg, M.~D. \& Petersen, M.~S. 2021, Monthly Notices of the Royal
  Astronomical Society, 501, 5408

\bibitem[{{Widmark} {et~al.}(2021){Widmark}, {Laporte}, {de Salas}, \&
  {Monari}}]{widmark2021weighing}
{Widmark}, A., {Laporte}, C.~F.~P., {de Salas}, P.~F., \& {Monari}, G. 2021,
  \aap, 653, A86

\bibitem[{Widrow {et~al.}(2014)Widrow, Barber, Chequers, \&
  Cheng}]{widrow2014bending}
Widrow, L.~M., Barber, J., Chequers, M.~H., \& Cheng, E. 2014, Monthly Notices
  of the Royal Astronomical Society, 440, 1971

\bibitem[{{Widrow} {et~al.}(2012){Widrow}, {Gardner}, {Yanny}, {Dodelson}, \&
  {Chen}}]{widrow2012galactoseismology}
{Widrow}, L.~M., {Gardner}, S., {Yanny}, B., {Dodelson}, S., \& {Chen}, H.-Y.
  2012, \apjl, 750, L41

\bibitem[{{Williams} {et~al.}(2013){Williams}, {Steinmetz}, {Binney},
  {Siebert}, {Enke}, {Famaey}, {Minchev}, {de Jong}, {Boeche}, {Freeman},
  {Bienaym{\'e}}, {Bland-Hawthorn}, {Gibson}, {Gilmore}, {Grebel}, {Helmi},
  {Kordopatis}, {Munari}, {Navarro}, {Parker}, {Reid}, {Seabroke}, {Sharma},
  {Siviero}, {Watson}, {Wyse}, \& {Zwitter}}]{williams2013wobbly}
{Williams}, M.~E.~K., {Steinmetz}, M., {Binney}, J., {et~al.} 2013, \mnras,
  436, 101

\end{thebibliography}

%\begin{thebibliography}{99}
%\end{thebibliography}

%%%%%%%%%%%%%%%%%%%%%%%%%%%%%%%%%%%%%%%%%%%%%%%%%%

%%%%%%%%%%%%%%%%% APPENDICES %%%%%%%%%%%%%%%%%%%%%

\begin{appendix}
\section{Misalignment with Z=0 plane}
\label{sec:misalignment_appendix}
\begin{figure}
	\includegraphics[width=1\columnwidth]{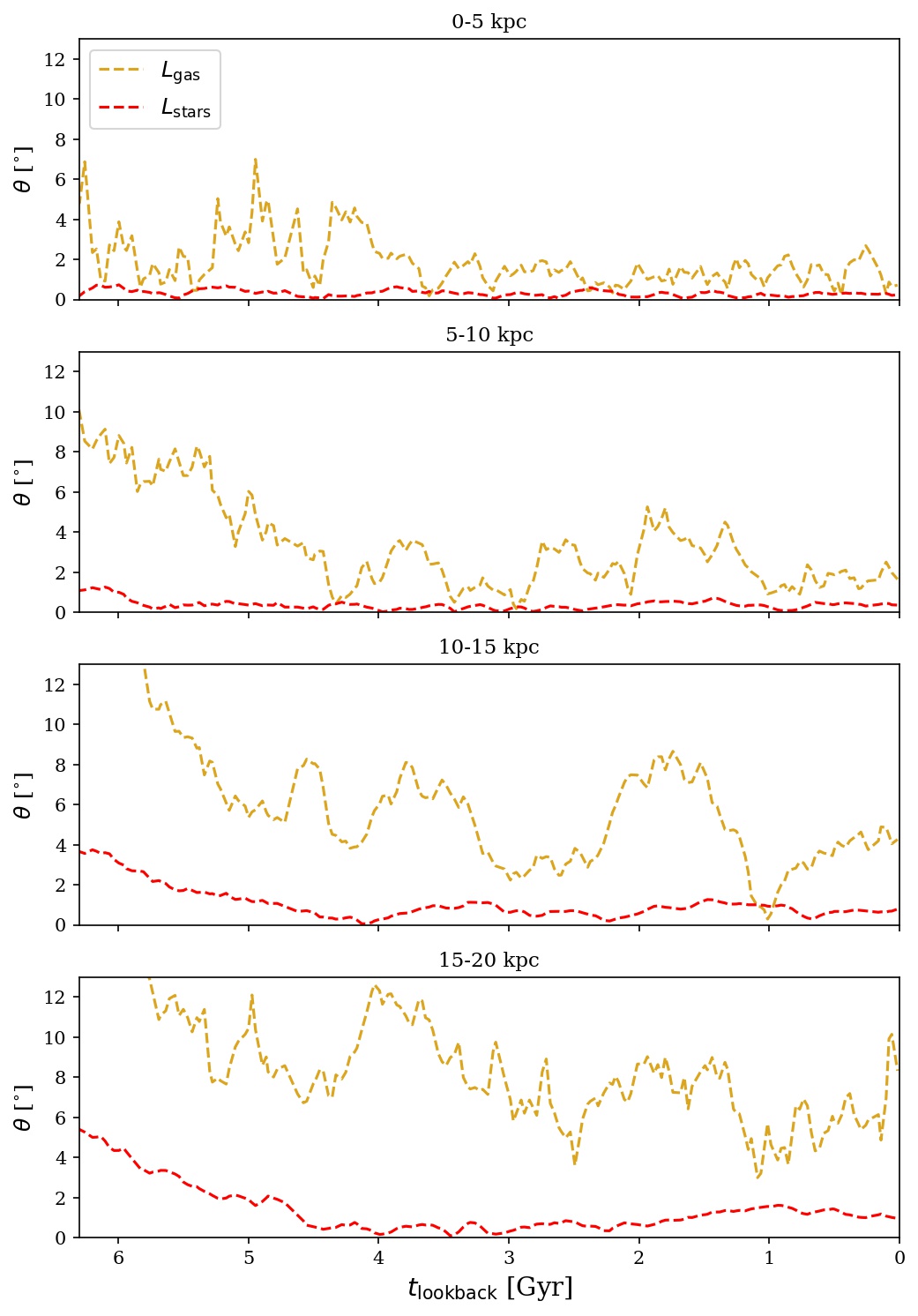}
    \caption{$L_{Z}$ misalignment of stars and gas in respect to the established $Z=0$ plane in different radial regions. From top to bottom, regions from 0 to 5 kpc, 5 to 10 kpc, 10 to 15 kpc and 15 to 20 kpc.}
    \label{fig:Lstars_vs_Lgas_ref_system}
\end{figure}

The definition of a galactic plane is not a trivial task. Here we show how some stellar particles and gas are misaligned with respect to our choice of the $Z=0$ plane,  which is aligned with stellar particles at $R<7$ kpc. In Fig. \ref{fig:Lstars_vs_Lgas_ref_system} we show this misalignment, $\theta$, for annular regions of 5 kpc of width from inner radii (top panel) to outer radii(bottom panel). In this reference system, those stellar particles at $R<5$ kpc (top panel) are well aligned with the plane, with generally less than 1$^{\circ}$ of misalignment. At the external radii, both stars and gas are less aligned with this plane, especially in the first gigayear of the period studied, due to the disk being highly perturbed at these times.

\section{Density and breathing modes}
\label{sec:density_breathing_relation}

\begin{figure*}
	\includegraphics[width=1\textwidth]{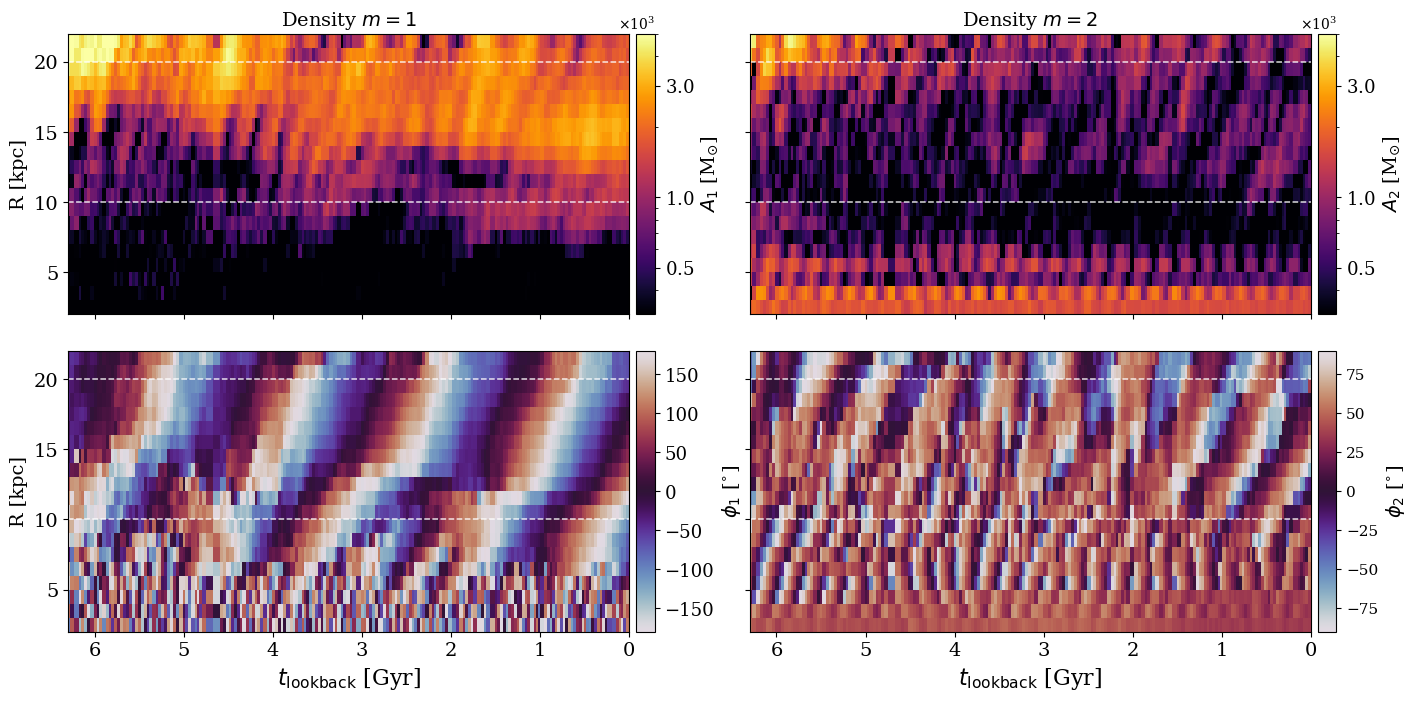}
    \caption{Similar to Fig.~\ref{fig:Bending_fourier} but for the stellar mass.}
    \label{fig:density_fourier}
\end{figure*}

In this section, we apply the same algorithms described in the main article to study the density structures of the disk. In this case, the weights in Eq. \ref{eq:fourier} are the mass of each stellar particle. The Fourier decomposition is shown in Fig.~\ref{fig:density_fourier}. In this figure, in $m=1$, we observe a retrograde, one-armed spiral, whose phase (bottom left panel) becomes more coherent from 4 Gyr on, and its amplitude (top left panel) appears to be higher at radii $> 7$ kpc. 
%The most important effect we observe at 4 Gyr is an increment in the amplitude of the density mode $m=1$ (a one-armed spiral structure that rotates counterclockwise, i.e. retrograde with respect to the stellar disk rotation). Since the increase of this first mode in density is slow, it is very difficult to ascertain when it was triggered. %It could be that the disk is still perturbed from the interaction with the dark satellite DM2512 at 5 Gyr approximately. 

%---link with density mode discussion. Origin? Arania? Dark Satellites?
The origin of this density mode is unclear. At external radii, this mode seems to exist since the beginning of our study, although at certain times (as discussed in previous sections), its amplitude seems to be higher. At radii where this amplitude is lower, the phase also becomes less clear and noisy. However, since its increases and decreases are not sudden, we do not see a clear link between this mode with other phenomena such as the passing of satellites. %Several studies have shown that perturbations of many different natures can generate a set of density pattern phenomena, such as density waves \citep{Bland-Hawthorn2021}. This may be the case for this model, where the enhancements of this density model roughly coincide with the passings of the satellite Arania, whose orbit is retrograde with respect to the direction of the rotation of stars. However, this link is not clear, since at the time of these pericenters, other satellites (such as the dark satellites crossing the disk at 4.1 Gyr) are also interacting with the disk. 
Our hypothesis is that, since this structure has an important presence throughout all timespan studied it is unlikely to be triggered by a single event, but instead by the structural density characteristics of the galactic system.
 %We also see a slight increase of the amplitude of density mode $m=2$ between 4 and 3 Gyr (panel H of Fig.~\ref{fig:1D_variables}) which may correspond to enhanced spiral arms.
In regard to the second Fourier mode (right panels), we see a difference between the inner and outer parts. The amplitude of the inner parts ($<$4 kpc) is dominated by the central oval, and from 4 to 7 kpc  periodic bisymmetric arms dominate.% which have higher Fourier amplitude when they reach the extremes of the central oval. 
The intermediate and outer parts are at times dominated by the amplitude of mode $m=2$, which could be related to spiral arms. In general, this mode propagates in the direction and periodicity of the rotation

\begin{figure*}
	\includegraphics[width=1\textwidth]{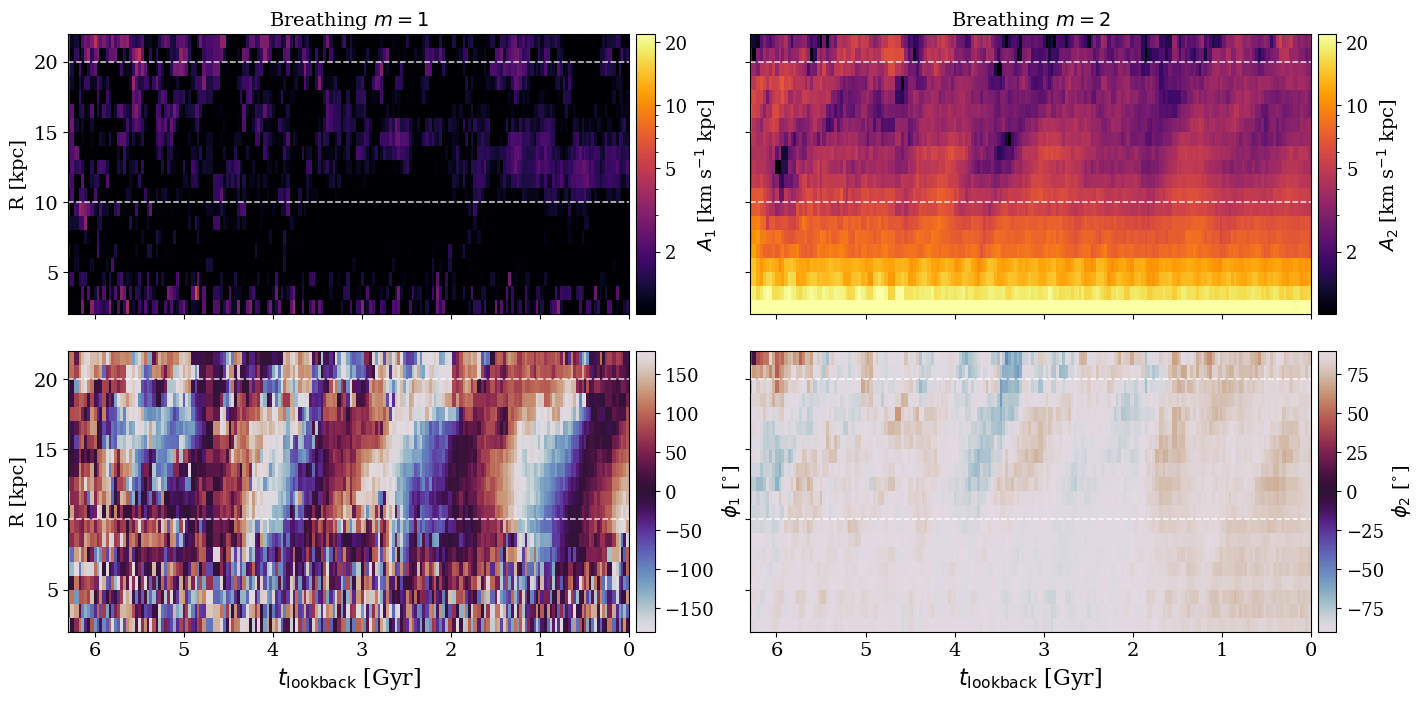}
    \caption{Similar to  Fig.~\ref{fig:Bending_fourier} but for the breathing modes.}
    \label{fig:Breathing_fourier}
\end{figure*}

In regards to the breathing modes in Fig.~\ref{fig:Breathing_fourier}, we observe that the amplitude is dominated by the Fourier mode $m=2$, which stays approximately constant between 0 and 7 kpc of radius and increases at certain times in the intermediate and outer regions of the disk. Its phase shows a small oscillation over time, but overall this angle coincides with the position of the disk's central oval, described in Sec. \ref{sec:garrotxa}. This is compatible with previous studies, where this breathing behavior is linked to the bar \citep{widrow2014bending, faure2014radial,monari2015vertical,  carrillo2018milky}.

Although weaker and noisier, the breathing mode $m=1$ becomes more noticeable (top left panel) and with a more clearly defined phase (bottom panel) from the period of 2 to 0 Gyr.  %At radii from 10 to 15 kpc we observe that the amplitude of the first mode in density and bending increase and correlate over the next gigayear, as well as with the dark matter acceleration at these regions. The breathing $m=1$ also increases, although very slightly above the noise. 
The studies by \cite{widrow2014bending,carrillo2018milky, hunt2021resolving} and \cite{kumar2022excitation} link the apparition of breathing modes in the disk with fast interaction with satellites. However, at these times, we do not have a fast (in terms of $V_{Z}$) crossing, but instead, the satellite Escarabajo has slow interaction and reaches its pericentre at higher $Z$. Thus, the increase in breathing mode as a direct consequence of the interaction with Escarabajo is improbable.
%The studies mentioned above do not establish a clear limit in which the breathing mode can appear. However, the increase in breathing mode as a direct consequence of the interaction with Escarabajo is improbable. 

%However, the increase in this breathing mode $m=1$ coincides with the increase in the amplitude of the one-armed density pattern, which could induce this kind of effect, since a concentration of mass could influence the vertical distribution of $V_{Z}$ of the particles entering and leaving the region with more density. The phase of this breathing mode $m=1$  (bottom panel) shows that this mode is slow and retrograde compared to the direction of rotation. This further supports that it may be related to the retrograde spiral density pattern.

\section{Sub-halos and dwarf galaxies as satellites}
\label{sec:satellites_appendix}
\begin{figure}
	\includegraphics[width=1\columnwidth]{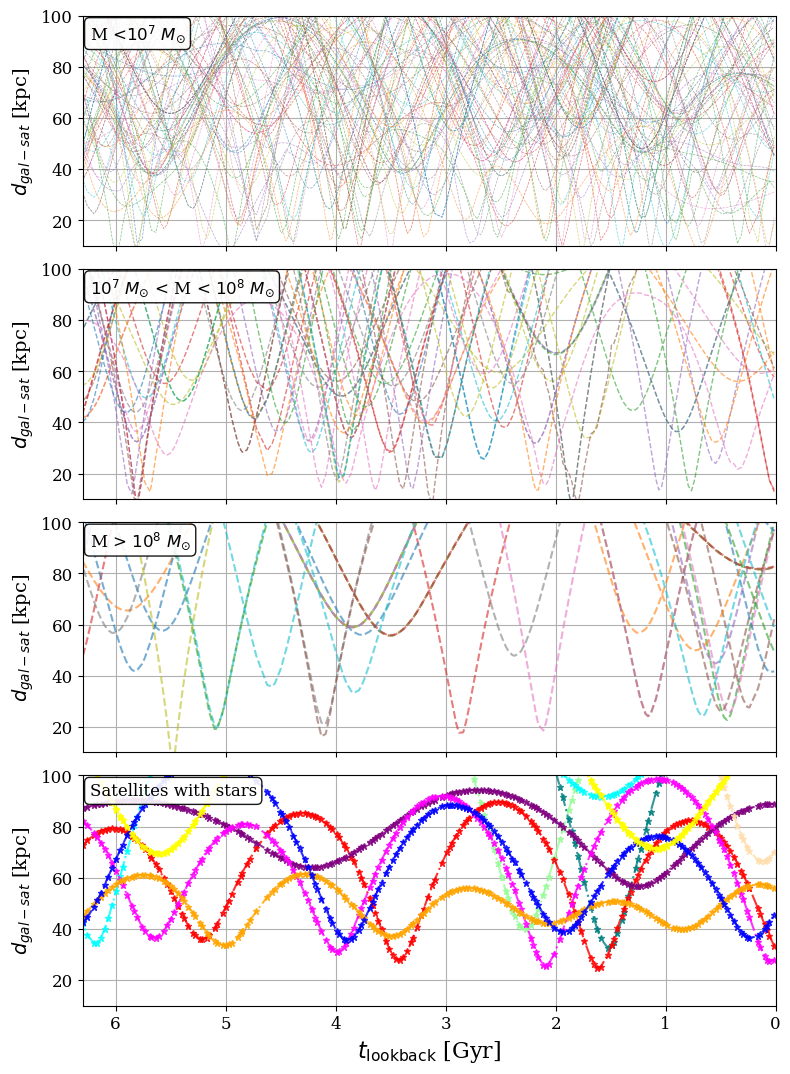}
    \caption{Distance of satellites to the centre of the galaxy. The top panel shows dark satellites of $10^{6}$ $M_{\odot}$ < M <$10^{7}$ $M_{\odot}$. The second panel are satellites $10^{7}$ $M_{\odot}$ < M <$10^{8}$ $M_{\odot}$. Third one  M > $10^{8}$ $M_{\odot}$. The last panel show all satellites with stellar mass.}
    \label{fig:satellites_by_mass}
\end{figure}

In this section, we show the pericentric moments of all satellites with $>10^{6}$ $M_{\odot}$ computed with the Rockstar halo finder. In Fig. \ref{fig:satellites_by_mass} we show the distance of the satellites to the center of the galaxy from less massive (top panel) to more massive ones (third panel). The bottom panel shows those galaxies with stellar particles.

\section{Correlations}
\label{sec:correlations_appendix}
In this work, we have calculated the accelerations of the dark matter particles and gas, which we have characterized in different regions using the Fourier mode $m=1$, which is dominant in both cases at most times. We have used the mean amplitude of said mode in regions from 0 to 10 kpc and 10 to 20 kpc. The same has been made with the variables tracing the vertical behavior of the disk, namely bending, breathing, and density modes. In this case, however, we also consider Fourier modes $m=2$ for bending, breathing, and density. 

\begin{figure}
	\includegraphics[width=1\columnwidth]{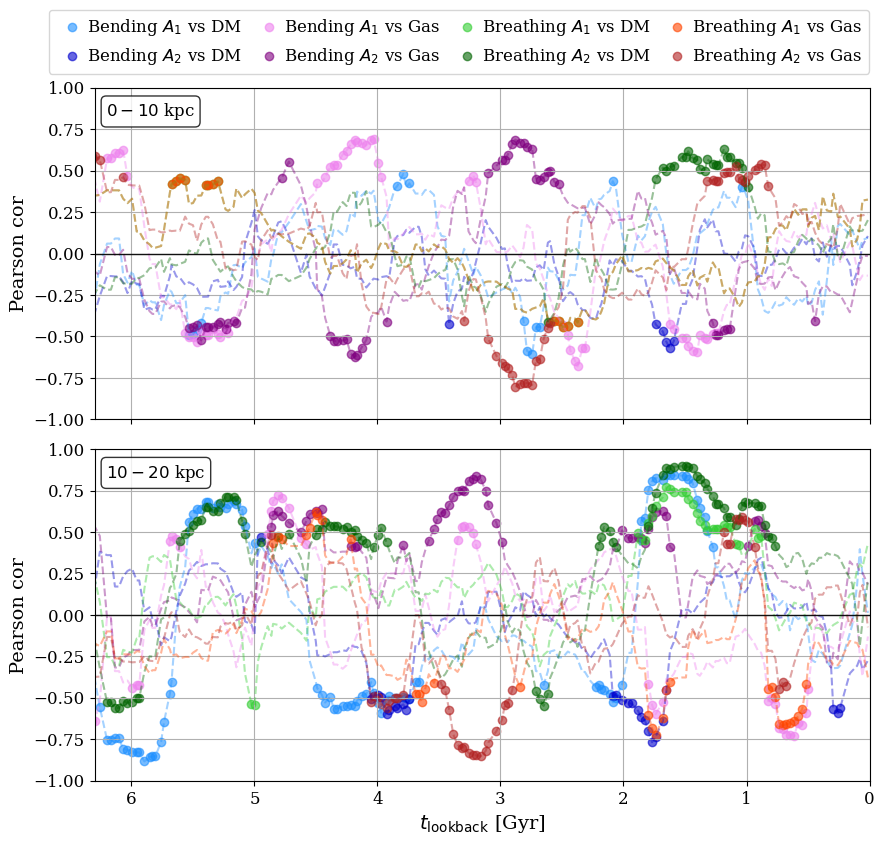}
    \caption{Pearson's correlation coefficient in moving windows of 800 Myr of width for bending and breathing modes vs accelerations of gas and dark matter. We mark as circles the moments where Pearson's coefficient fulfils $|r|>0.4$ and $p-value < 0.05$}
    \label{fig:cor}
\end{figure}

\begin{figure}
	\includegraphics[width=1\columnwidth]{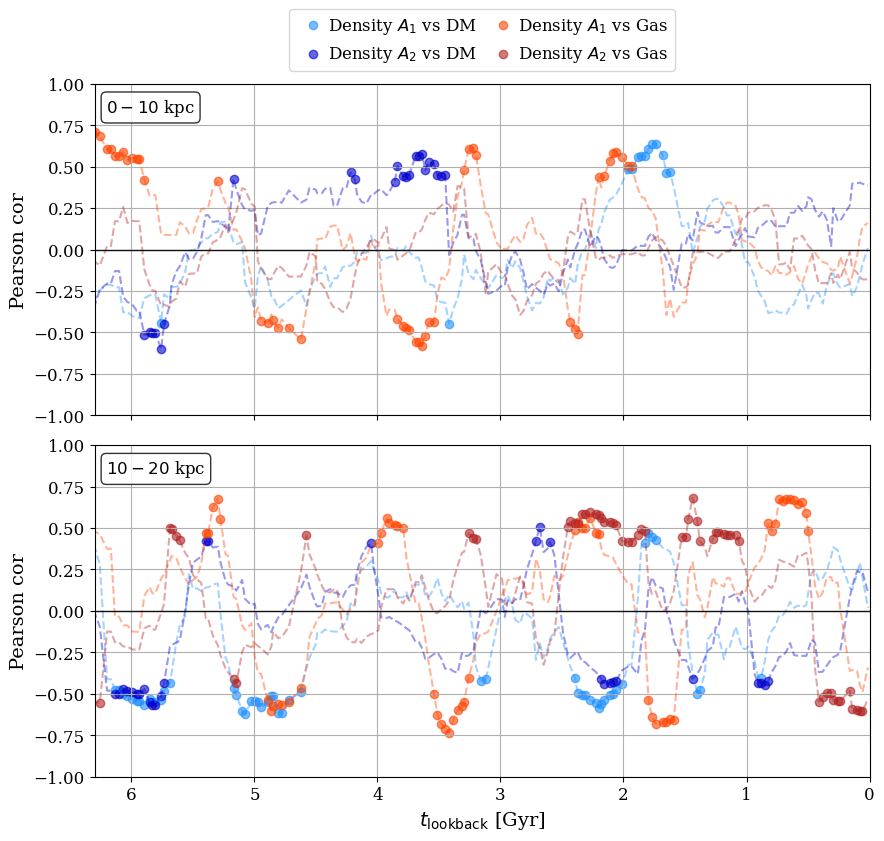}
    \caption{Same as Fig \ref{fig:cor} but with variables of density vs gas and dark matter acceleration's $m=1$.}
    \label{fig:cor2}
\end{figure}

\begin{figure}
	\includegraphics[width=1\columnwidth]{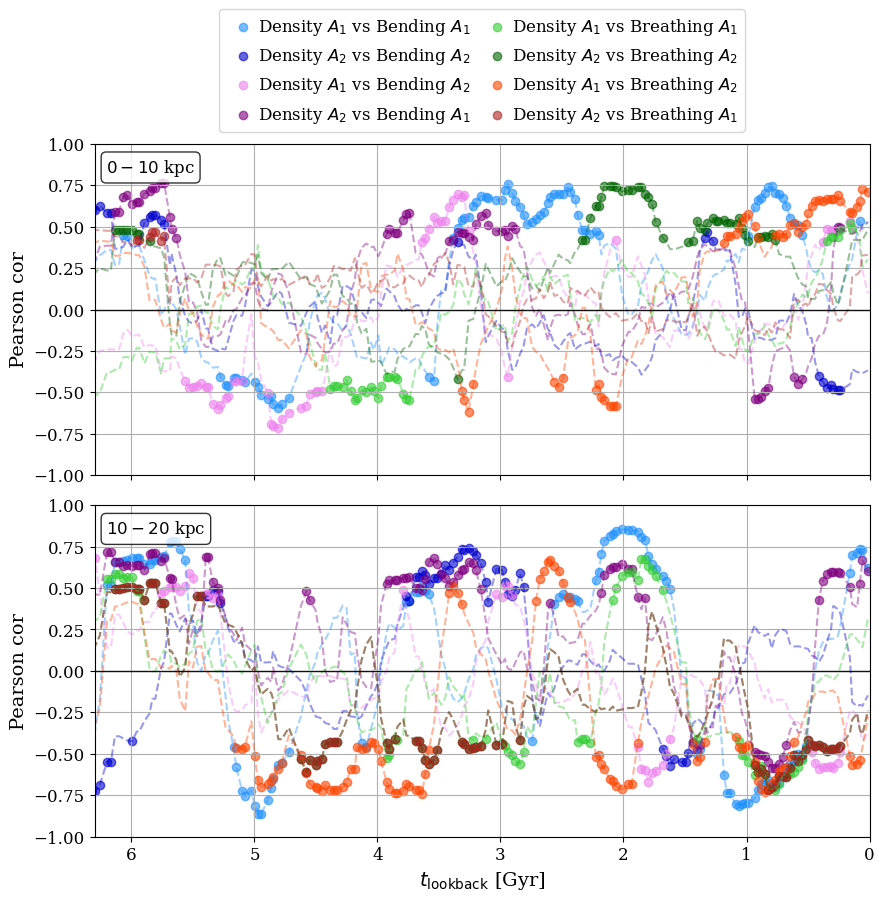}
    \caption{Same as Fig \ref{fig:cor} but with variables of density vs bending and breathing modes.}
    \label{fig:cor3}
\end{figure}

To begin to understand the relationships between all these variables at all times we define a moving window of 800 Myr of width, moving along lookback time. At each snapshot, we define a window spanning from the previous 400 Myr to 400 Myr later. We calculate Pearson's correlation coefficient $r$ and its p-value at each window. Since the pairs of variables are numerous, we present the results in three different figures. In Figs. \ref{fig:cor}, \ref{fig:cor2}, and \ref{fig:cor3} we present the correlation coefficient (lines), and the most notable correlations (where  $|r|>0.4$ and p-value $< 0.05$ circles).

These figures serve as a visual aid to identify in which moments different variables increase or decrease in relation to each other, and, combined with Fig.~\ref{fig:1D_variables}, can help to better understand the causes and effects at play at the different time intervals discussed in Sec. \ref{sec:results}. However, due to the small number of snapshots per window (average of 23 snapshots), we cannot extract accurate and definite conclusions based on these correlations on their own.

In the three figures we observe that the first gigayear of the studied period presents high correlations between all variables due to the highly perturbed state of the galaxy and the subsequent relaxation over the next gigayear. We have a special interest in the correlations with the bending $m=1$ at external radius starting at 2 Gyr, since it is the higher bending mode we find over time and radius. In the bottom panel of Fig. \ref{fig:cor} we find a high correlation between this bending mode and the mode $m=1$ of $a_{Z}$ applied by dark matter. We also find breathing modes being excited at these times, which also correlate with the acceleration by dark matter at these times.

From 2.2 Gyr until 1.6 Gyr the density mode $m=1$ correlates with the acceleration by dark matter and gas  at more internal radii between 0 and 10 kpc (top panel of Fig.~\ref{fig:cor2}, light blue and orange dots, respectively). At external radii (bottom panel of said figure) there is not a clear tendency. At these times and radii, in the top panel of Fig.~\ref{fig:cor3}, we also see a correlation between the density mode $m=2$ and the breathing mode $m=2$ (dark green dots) and between the modes $m=1$ of density and bending (light blue dots), which are also seen between 10 and 20 kpc.

%%%%%%%%%%%%%%%%%%%%%%%%%%%%%%%%%%%%%%%%%%%%%%%%%%
\end{appendix}

% Don't change these lines
%\bsp	% typesetting comment
%\label{lastpage}

\end{document}